\documentclass[sigconf]{acmart}
\AtBeginDocument{%
  }

\setcopyright{acmlicensed}

\copyrightyear{2025}
\acmYear{2025}
\setcopyright{cc}
\setcctype{by}
\acmConference[FAccT '25]{The 2025 ACM Conference on Fairness,
Accountability, and Transparency}{June 23--26, 2025}{Athens, Greece}
\acmBooktitle{The 2025 ACM Conference on Fairness, Accountability, and
Transparency (FAccT '25), June 23--26, 2025, Athens,
Greece}\acmDOI{10.1145/3715275.3732030}
\acmISBN{979-8-4007-1482-5/2025/06}

\usepackage{xcolor}
\usepackage{graphicx} 
\usepackage{booktabs} 
\usepackage{xfrac}
\usepackage{wrapfig} 
\usepackage{multirow}
\usepackage{stfloats}
\usepackage{enumitem}
\usepackage{tabularx}
\usepackage{subcaption}
\usepackage{epigraph}

\newcommand{\fixedbox}[3]{%
  \hspace{1pt}\raisebox{0.1ex}{
    \colorbox{#1}{%
      \hspace{-1pt}
      \rule[-0.1ex]{0pt}{1.5ex}
      \textcolor{#2}{\textsf{\small #3}}%
      \hspace{-1pt}
    }%
  }%
}

\newcommand{\agonisticbox}{\fixedbox{yellow!20}{yellow!80!black}{\textsf{\textsc{Agonistic}}}~}
\newcommand{\baselinebox}{\fixedbox{gray!10}{gray!90}{\textsf{\textsc{Baseline}}}~}
\newcommand{\reformulativebox}{\fixedbox{red!10}{red!80!black}{\textsf{\textsc{Reformulative}}}~}
\newcommand{\diversebox}{\fixedbox{blue!10}{blue!80!black}{\textsf{\textsc{Diverse}}~}}

\newcommand{\agonistic}{\textsf{\textsc{Agonistic}}}
\newcommand{\baseline}{\textsf{\textsc{Baseline}}}
\newcommand{\reformulative}{\textsf{\textsc{Reformulative}}}
\newcommand{\diverse}{\textsf{\textsc{Diverse}}}

\newcommand{\direct}{\textsf{Direct}}
\newcommand{\reminder}{\textsf{Reminder}}
\newcommand{\expansion}{\textsf{Expansion}}
\newcommand{\challenge}{\textsf{Challenge}}

\newcommand{\realism}{\textsf{Realism}}
\newcommand{\familiarity}{\textsf{Familiarity}}
\newcommand{\diversity}{\textsf{Diversity}}
\newcommand{\aesthetics}{\textsf{Aesthetics}}

\newcommand{\pref}[1]{\textsf{P#1}}

\begin{document}

\title{Agonistic Image Generation: \\Unsettling the Hegemony of Intention}

\author{Andrew Shaw}
\authornote{Both authors contributed equally to this research.}
\orcid{0009-0007-4579-718X}
\author{Andre Ye}
\authornotemark[1]
\orcid{0000-0003-4936-7914}
\affiliation{%
  \institution{University of Washington}
  \city{Seattle}
  \state{Washington}
  \country{USA}
}
\email{{shawan, andreye}@uw.edu}

\author{Ranjay Krishna}
\affiliation{%
  \institution{University of Washington,\\Allen Institute for AI}
  \city{Seattle}
  \state{Washington}
  \country{USA}}
\email{ranjay@cs.uw.edu}

\author{Amy X. Zhang}
\affiliation{%
  \institution{University of Washington,\\Allen Institute for AI}
  \city{Seattle}
  \state{Washington}
  \country{USA}
}
\email{axz@cs.uw.edu}

\renewcommand{\shortauthors}{Shaw $\&$ Ye et al.}

\begin{abstract}

Current image generation tools largely follow an intention-centric paradigm that aims to actualize user intentions but neglects the sociopolitical conversations that these intentions are embedded in.
As these tools become cornerstones of the media landscape, however, it is increasingly evident that sociopolitical conflicts over visual representation are inescapable parts of image generation.
For instance, in March 2024, Google's Gemini faced criticism for inappropriately injecting demographic diversity into user prompts.
Although Gemini challenged user intentions, its top-down imposition of a standard for ``diversity'' ultimately proved counterproductive.
In this paper, we present an alternative approach: an image generation interface designed to embrace open negotiation along the sociopolitical dimensions of image creation.
Grounded in agonistic pluralism (from the Greek \textit{agon}, meaning struggle), our interface actively engages users with competing visual interpretations of their prompts.
Through a lab study with 29 participants, we evaluate our agonistic interface on its ability to facilitate reflection, or engagement with other perspectives that challenges dominant assumptions.
We compare it to three existing paradigms:
a baseline interface that emulates current image generation tools,
a Gemini-style interface that produces ``diverse'' images, 
and an intention-centric interface that suggests aestheticized prompt refinements.
We find that the agonistic interface enhances reflection across multiple measures, but also that reflection depends on authentically grounding diversity in relevant political contexts.
Our results suggest that diversity and user intention need not be treated as inherently opposing values.
Instead, interfaces can productively navigate tensions between competing perspectives, enabling users to engage with and evolve their intentions in meaningful ways.

\end{abstract}



\maketitle

\epigraph{``\textit{One does not become enlightened by imagining figures of light, but by making the darkness conscious.}'' -- Carl Jung}{}

\section{Introduction: The Hegemony of Intention}
\label{intro}

\textit{What does ``a Founding Father'' look like? What does ``Jesus'' look like? What about ``a migrant,'' ``a doctor,'' ``World War II''?}
Today, the ``answers'' to these questions increasingly draw from AI-generated images proliferating in our digital media ecosystem \cite{jingnan2024ai, diresta2024spammers, herrieetal2024democratization}.
The politics of image generation were underscored in March 2024 when Google's Gemini model came under criticism for generating images with inaccurate racial and gender diversity, such as depictions of ``Founding Fathers'' as women of color \cite{milmoandkern2024gemini}.
While many found these historically inaccurate images of Founding Fathers to be problematic, public acclaim of works like the Broadway play \textit{Hamilton}, which used actors of color to portray Founding Fathers, implies that depicting historical figures in ``inaccurate'' ways can nonetheless carry valuable artistic and political messages.
These examples show how the production and consumption of visual meanings is a complex \textit{sociopolitical} endeavor, and the values embedded in image generation tools shape how we engage with our social and political world.
When people consider their mental image of ``a Founding Father,'' (or ``Jesus'', ``a migrant'', etc.) their preconceptions are embedded in dynamic ongoing conversations with others about who a Founding Father is and---just as importantly---is not.
Image generation interfaces are therefore part of these complex conversations and should be designed with this in mind.

Despite the political character of image generation, the dominant paradigm behind many image generation tools remains what \citet{sarkar2024intention} calls an ``intention is all you need'' paradigm that prioritizes the \textbf{actualization of intention}---aiming for users to ``see what they intend.''
While this paradigm may make users happy, it also obscures the conflictual nature of visual representation---that users' mental image of a subject may only be a part of a larger conversation with others have different mental images of that subject.
When users do not engage with this discourse, their ideas may be unchallenged, detached, and polarized~\cite{cai2024antagonisticai}.

The Gemini controversy certainly challenged this ``hegemony of intention'' by engaging users with political controversies over diversity and representation.
However, it did so in a way that caused significant public backlash, inciting ``\textit{an explosion of negative commentary from figures such as Elon Musk}'' that prompted Google to rescind its changes \cite{milmoandkern2024gemini}.
The swift pushback suggests that top-down interventions may be ill-suited to the dynamic political landscape of image generation.
In short, the Gemini case demonstrates that interventions which do not \textit{actively engage} users in discourse but rather impose on their intent can come across as patronizing infringements on user agency. 

In this work, we build and evaluate an image generation interface that recenters the political context of image generation by encouraging users to actively engage with a wide range of conflicting interpretations \textit{on their own terms}. 
Our design draws from \textbf{agonistic pluralism} (from the Greek \textit{agon} meaning ``struggle''), a political philosophy emphasizing the importance of healthy democratic conflict. 
Given a user prompt, our interface \agonistic{} highlights ongoing political controversies over the user prompt by collecting and presenting a range of conflicting interpretations to the user (Figure~\ref{fig:interface-screenshots}).
Only after engaging with different interpretations and selecting one does a user then receive image generations.

A critical part of agonistic pluralism is \textit{reflection}---a process that involves engaging with competing viewpoints and challenging dominant assumptions. 
For this reason, we conduct a comparative lab evaluation of our interface with 29 participants on its ability to foster productive critical reflection.
We build three other paradigmatic interfaces for comparison: 
\baseline: produces images generated from the user's prompt with no further interaction;
\diverse: rewrites the user's prompt to be more diverse before image generation;
and \reformulative: provides prompt reformulations to produce more aesthetic and preferable images, inspired by existing work in prompt reformulation.
Across several measures of reflection, we find that \agonistic{} induces the highest reflection out of any interface (\S\ref{how-reflection}), followed by \reformulative, then \diverse.
For instance, when participants went to add images to a collage, they described the image as an expansion of their original intention 21\% of the time while using \agonistic, but only 8\%, 7\%, and 3\% of the time for \reformulative, \diverse, and \baseline, respectively.
Our findings confirm that participants are likely to reject Gemini-style interventions due to a lack of factuality and authenticity, while \agonistic{} more authentically expresses diversity by situating the user prompt in relevant sociopolitical discourse.
Thus, getting users to acknowledge and prioritize diversity requires situating diversity in authentic political controversies, departing from existing interventions that either seek to impose diversity for its own sake or constrain diversity with factuality.
Instead of imposing top-down interventions onto users, AI tools can instead create space for open-ended democratic contestation over values like user intention, factuality, and diversity. 
Our findings highlight the importance of bringing HCI work to bear on AI ethics and provide more general design elements for building reflective AI systems.
\section{Background and Related Work}
\label{background}

\subsection{Images as Sociopolitical Artifacts}
\label{discursive-approaches}

Philosophers of semiotics have long argued that signs (symbols like words or images representing something other than themselves) can be ``performed'' to demonstrate membership within social categories \cite{butler1990gender, hall1990cultural, hall1997representation} and invoked to effect changes in the world \cite{austin1956performative, austin1962how}. 
In other words, signs do not just represent, but also play active roles in social life. 
Likewise, philosophers of history have challenged concepts of ``objective'' truth by calling attention to complex networks of competing narratives, power relations, and institutional practices~\cite{certeau1988writing, white1973metahistory}---what Michel Foucault calls ``discourse''~\cite{foucault1972archaeology}---that shape historical record and study.
Thus, historical questions of ``how things were really like'' are far from straightforward because appealing to historical accuracy also requires considering how discourse shapes the historical record itself.
Media studies scholars have further elaborated upon these foundational insights in the context of visual media.
Images are themselves political not only in the sense that interpretation of them is conditioned by cultural context \cite{barthes1981camera}, but also in the sense that the act of creating or synthesizing images always involves ``\textit{imposing standards on [one's] subject}''~\cite{sontag1977on}.
These works suggest that all image production and consumption---including AI image generation---is implicated in social, political, and ethical issues~\cite{mulvey1975visual}: ``\textit{individuals implicitly engage in ongoing struggles over visual dominance and its articulation with social formations}''~\cite{herrieetal2024democratization}.
Our aim in this paper is to build an image generation interface which engages users in this complexity of image generation.

\subsection{Agonistic Pluralism: A Political Framework for Image Generation}
\label{agonistic-democracy}

Given the political nature of image generation, what kind of political structures should be used to deal with issues of diversity and representation?
Some commentators have found the problem with the Gemini case to be that ``\textit{tuning to ensure that Gemini showed a range of people failed to account for cases that should clearly not show a range}'' \cite{milmoandkern2024gemini}. 
However, scholars like \citet{melamed14diversity} argue that too often, categories of ``good'' and ``bad'' diversity only end up legitimizing existing relations of domination---``good'' forms of diversity are those that comport with hegemonic social narratives, whereas ``bad'' forms of diversity are those that challenge them. An alternative approach is to abandon the expectation of a clear division between good and bad forms of diversity, and instead embrace a radically democratic approach that recognizes an arena of conflicting views that challenge dominant assumptions. For instance, as alluded to in \S\ref{intro}, the differing social responses to the Gemini controversy and \textit{Hamilton} can be thought of as part of a larger conflict over representations of American culture and history.

To this effect, \citet{herrieetal2024democratization} propose agonistic pluralism as a particularly apt theory of democracy for negotiating the political dimensions of image generation. 
First developed by \citet{mouffe2000democraticparadox}, agonistic pluralism is a philosophy that sees political life as being inescapably structured by conflict between competing views.
The task for democracy is thus not to eradicate such conflict, however, but to channel it into productive ends.
To do so, democracy should always allow for the possibility of political conflict by providing avenues to challenge hegemonic assumptions.
However, it should also work to construct political conflict in adversarial rather than antagonistic terms: in contrast to purely antagonistic struggle in which opponents view each other as enemies to be destroyed, Mouffe explains that agonistic struggle requires viewing opponents legitimate adversaries ``\textit{whose ideas we combat but whose right to defend those ideas we do not put into question}'' \cite{mouffe2000democraticparadox}.
Agonistic pluralism is therefore critical of top-down interventions like the Gemini case which impose a singular standard of visual representation and eliminate the possibility for users themselves to engage in broader political conflicts over visual prompts.

\subsection{Values in Image Generation Interface Design}
\label{current-work}

Many current works aim to help users produce ``better'' images without necessitating additional user interaction, invoking the value of \textbf{intention actualization}.
Prompt reformulation methods automatically rewrite prompts to produce more aesthetic or generally preferred~\cite{zhan2024capability, xu2024imagereward, datta-etal-2024-prompt, chen2024tailored} images, motivated by ``\textit{convey[ing] the user’s \textbf{intended ideas}}''\cite{mo2024dynamicpromptoptimizingtexttoimage} or ``\textit{preserving the 
\textbf{original user intentions}}''~\cite{hao2024optimizing}.
Other work builds interfaces to help users explore and iterate over prompts, invoking similar values:
``\textit{align the user’s creative \textbf{intentions} with the model’s generation'' }~\cite{wang2024promptcharm};
``\textit{align with [users'] \textbf{intended creative output}}''~\cite{brade2023promptify}.
This emphasis on actualizing user intentions, and its potential pitfalls, is emblematic of what \citet{sarkar2024intention} calls the ``intention is all you need'' paradigm. ``\textit{Contrary to the assumption that GenAI merely executes human intentions, it also shapes them [...] induc[ing] `mechanised convergence,' homogenising creative output, and reducing diversity in thought}.''
Interfaces that overly emphasize ease of use may be ``\textit{easy to learn at the expense of the power and complexity necessary to do hard but valuable work or learn uncomfortable truths}''~\cite{cai2024antagonisticai}.
In response to these critiques, our work explores how users might \textit{question} and \textit{revise} their intentions, for example by reflecting upon limiting assumptions. 

Another notable value in image generation is \textbf{diversity}, especially human diversity, which we focus on in this paper.
Image generation models have well-documented issues with diversity, often failing to appropriately represent demographic minorities and resorting to stereotypes~\citep[\textit{inter alia}]{ali2024demographic, wang-etal-2023-t2iat, wang2024newjob, chauhan2024identifying, naik2023t2i}.
Proposed solutions to these issues include finetuning the model with some diversity objective~\cite{miao2024training, esposito2023mitigatingstereotypicalbiasestext}, intervening on internal model states~\cite{friedrich2023fairdiffusioninstructingtexttoimage, zhang2023iti, zameshina2023diversediffusionenhancingimage, chuang2023debiasingvisionlanguagemodelsbiased}, or rewriting the prompt to emphasize diversity~\cite{microsoft_prompt_transformation}.
Our work takes a different, \textit{interactive} angle on diversity, aiming to engage users with competing visual interpretations of a subject rather than directly generating ``diverse'' images (see \S\ref{inauthentic-diversity}).

\subsection{Agonistic Design and Reflection in HCI}
\label{hai-reflection}

In contrast to the intention-centric paradigm, \citet{crawford2014agonistic} proposes an agonistic design paradigm which begins from the premise that ``\textit{complex, shifting negotiations are occurring between people, algorithms, and institutions, always acting in relation to each other.}'' By foregrounding such political negotiations, agonistic design is also similar to design approaches like adversarial design \cite{disalvo12adversarialdesign}, antagonistic design \cite{cai2024antagonisticai}, and other approaches that seek to disrupt rather than reproduce user intentions in human-computer interactions \cite{sarkar2024intention, herrieetal2024democratization}. 
Agonistic design also finds affinities in HCI work on designing for reflection, where reflection is commonly understood as ``\textit{intellectual and affective activities in which individuals engage to explore their experiences in order to lead to new understandings and appreciations}'' \cite{boudetal1985reflection}.
\citet{fleckandfitzpatrick2010reflectingonreflection} build on this definition with a five-level schema of reflection, in which higher levels of reflection involve increasingly deep engagement with other views and critical examination of biases. The highest level, critical reflection, can be understood as the result of a two-stage process in which users first encounter a discomforting experience, and then become critically aware of their own biases by working through discomforting feelings \cite{halbertandnathan2015criticalreflection}.
\citet{angelietal21agonisticgames} thus conceive of critical reflection as a key aim of agonistic pluralism and propose three tenets of agonistic design for video games to support critical reflection: confronting users with unsettling choices, encouraging engagement with multiple perspectives, and situating subject matter in a context that resonates with users. 
Nonetheless, art and culture remain an underexplored area of HCI research on reflection~\cite{bentvelzenetal2022revisitingreflection}, and to our knowledge no prior work has applied agonistic design to image generation tools.
\section{Four Image Generation Interfaces}
\label{paradigms-interfaces}

\begin{figure*}
    \centering
    \begin{subfigure}[b]{0.35\textwidth}
        \centering
        \includegraphics[height=6cm]{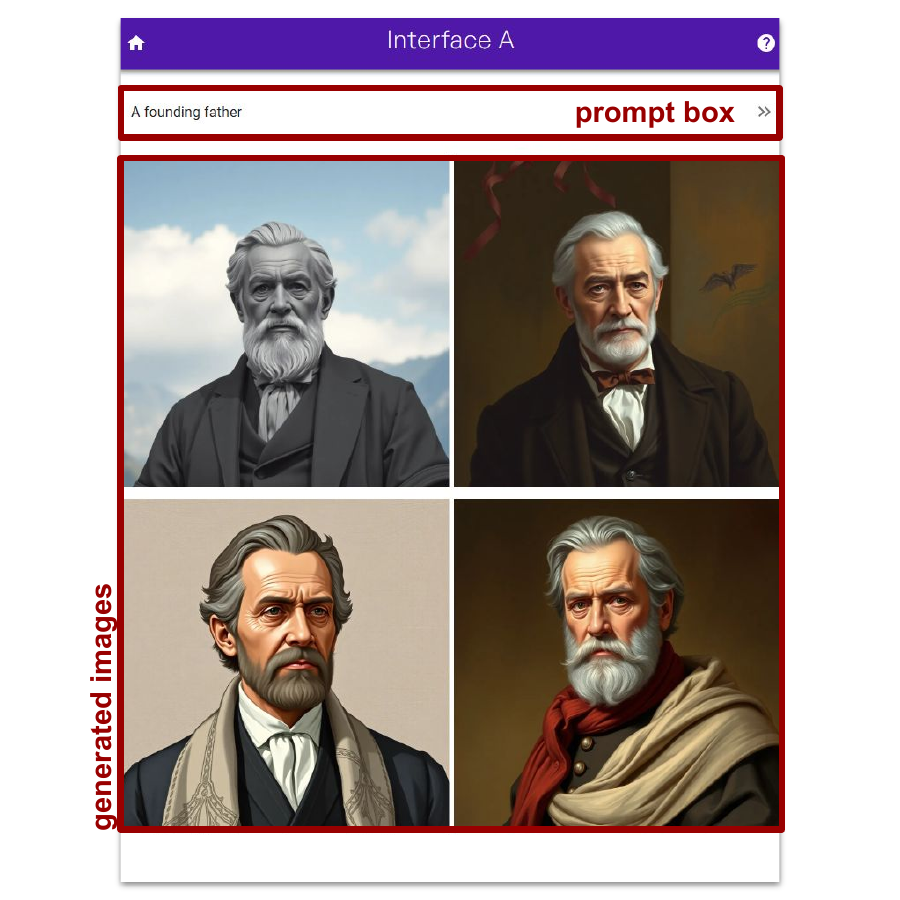}
        \phantomsubcaption
        \caption*{(a) \baselinebox}
        \label{fig:demo-baseline}
    \end{subfigure}
    \hfill
    \begin{subfigure}[b]{0.64\textwidth}
        \centering
        \includegraphics[height=6cm]{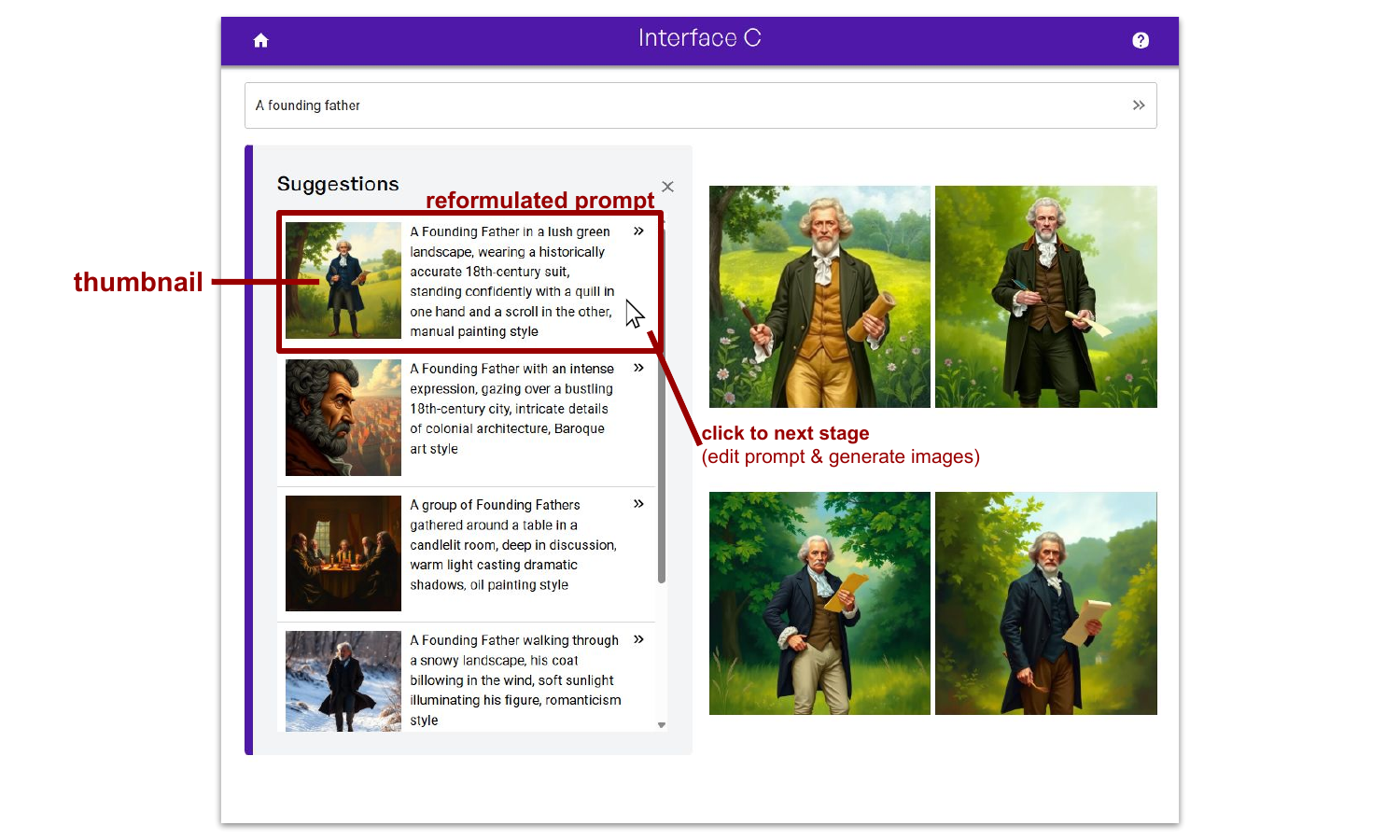}
        \phantomsubcaption
        \caption*{(c) \reformulativebox}
        \label{fig:demo-reformulative}
    \end{subfigure}
    \vspace{4mm}
    \begin{subfigure}[b]{0.35\textwidth}
        \centering
        \includegraphics[height=6cm]{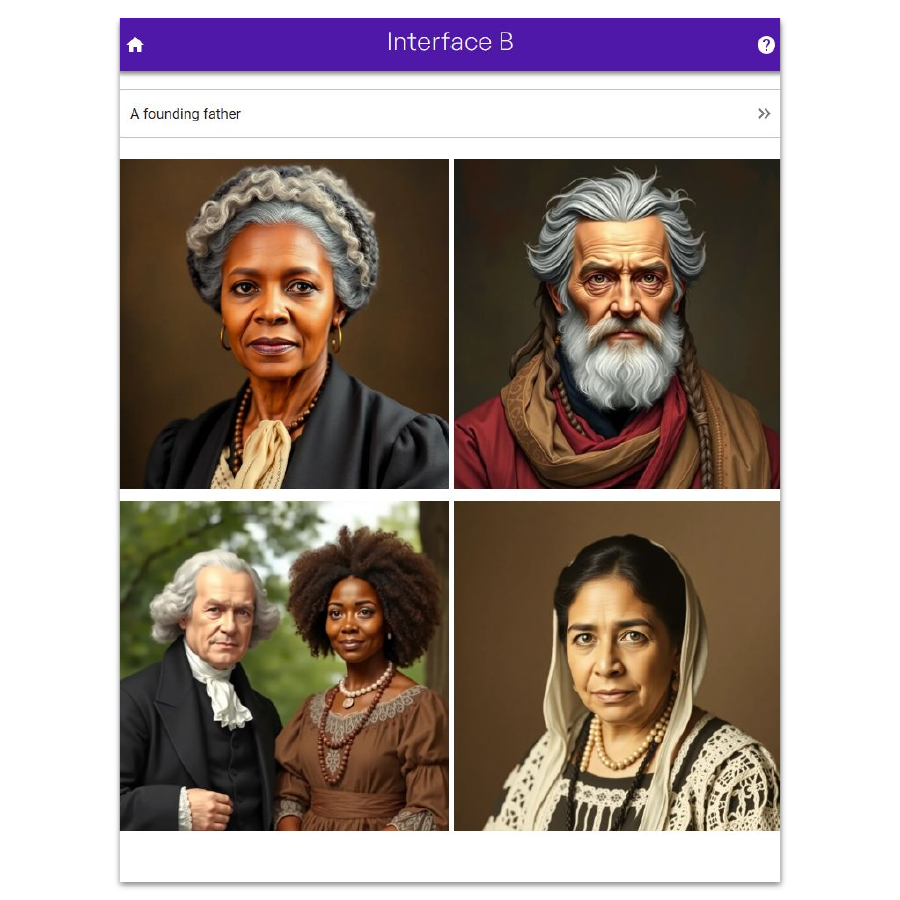}
        \phantomsubcaption
        \caption*{(b) \diversebox}
        \label{fig:demo-diverse}
    \end{subfigure}
    \hfill
    \begin{subfigure}[b]{0.64\textwidth}
        \centering
        \includegraphics[height=6cm]{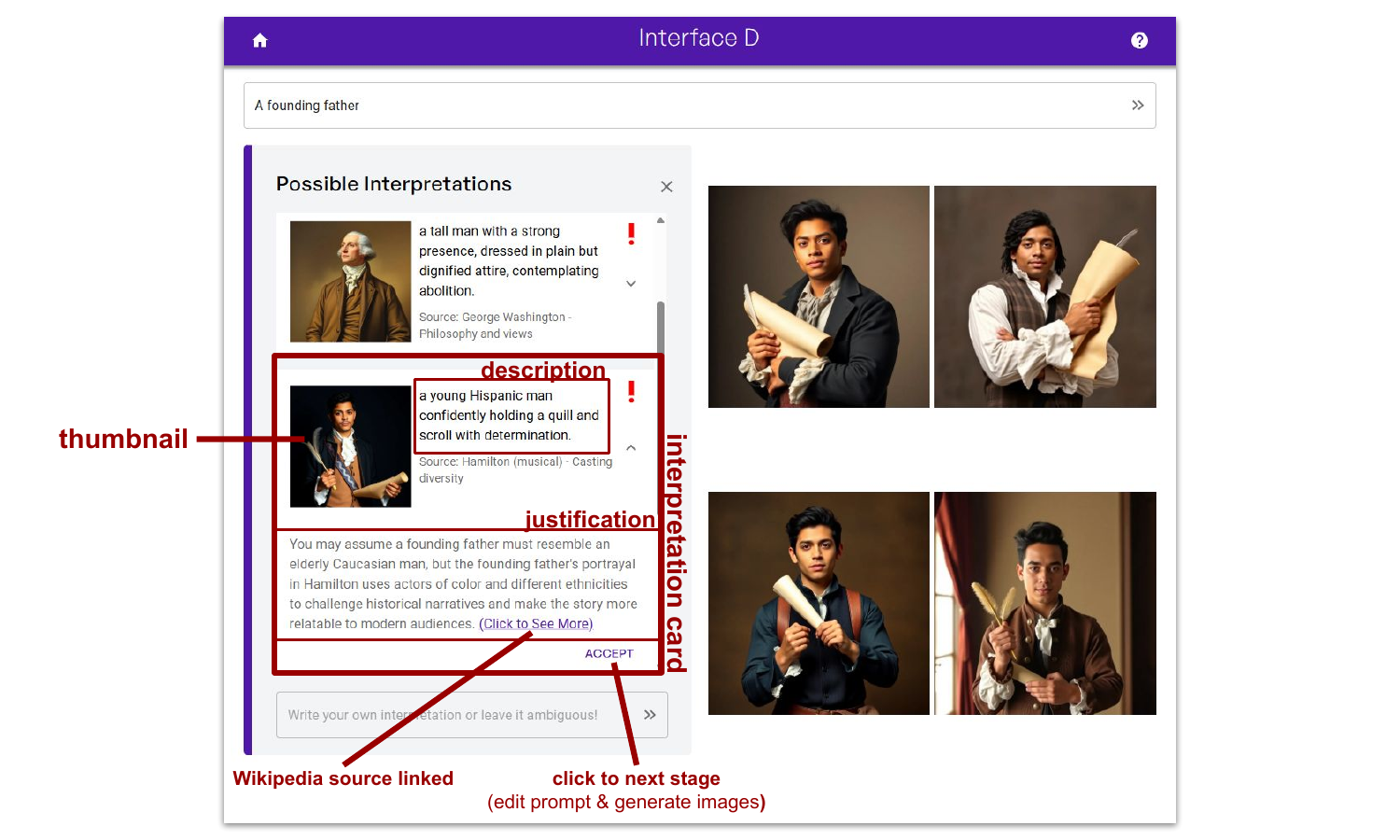}
        \phantomsubcaption
        \caption*{(d) \agonisticbox}
        \label{fig:demo-agonistic}
    \end{subfigure}
    \caption{Screenshots from each image generation interface evaluated in lab study. See \S\ref{paradigms-interfaces} for details. Note that \agonisticbox~depicts men of color as a possible interpretation of ``Founding Father,'' drawing upon the play \textit{Hamilton}. This interpretation is displayed among many others to engage users in discourse about what a Founding Father looks like.}
    \label{fig:interface-screenshots}
\end{figure*}

To holistically evaluate how agonistic design affects user reflection, we create a \baseline{}  interface, which accepts the user's prompt and outputs generated images with no further interaction, and three \textit{paradigmatic interfaces} each embodying a dominant approach or value: \diverse{} for diversity, \reformulative{} for intention actualization, and \agonistic{} for agonism.
In this section, we describe the key frontend and backend elements of each interface.

\noindent
\textbf{\baselinebox}~$\cdot$~
We design a \baseline{} interface as a control for the study. In this interface, users enter a prompt and are displayed four images generated from the prompt.
We use a lightweight open-source image generation model (Black Forest Labs, \verb|flux-schnell|) throughout all interfaces in the study, to avoid relying on interfaces that perform under-the-hood prompt rewriting (e.g., DALL-E API).

\noindent
\textbf{\diversebox}~$\cdot$~
The \diverse{} interface represents a paradigm of image generation that explicitly corrects for diversity issues by rewriting prompts in more diverse ways, as reported in the Gemini case. 
To create this interface, we use an allegedly leaked DALL-E prompt as reported by \citet{milmoandkern2024gemini}, which instructs the model to ``\textit{diversify depictions of ALL images with people to include DESCENT and GENDER for EACH person using direct term.}''
We instruct GPT-3.5 to rewrite the user prompt with the above instructions and use the rewritten prompt to generate four images that are then displayed to the user.
Although we recognize that the prompt may not match the exact approach used in the Gemini case, it anecdotally yields similar results to reported images for a variety of prompts.

\noindent
\textbf{\reformulativebox}~$\cdot$~
The \reformulative{} interface embodies the value of intention actualization and is inspired by \S\ref{current-work}; it supports users by producing prompt reformulations that add detail to the prompt (e.g., adding useful semantics or stylistic indicators) in ways that are likely to result in more aesthetic or preferred images.
This interface is similar to interfaces like Promptify, which allows users to iteratively refine their prompts based on a variety of AI-generated reformulations \cite{brade2023promptify}.
Similarly, \citet{skiltonandcardinal2024inclusivepromptengineering} propose using AI to generate a list of ``suggested descriptors'' that the user can then choose from and modify to create images that depart from stereotypical representations. 
We instruct GPT-4o to generate a diverse set of detail-adding reformulations of the user prompt with in-context examples from or inspired by Promptify.
To aid visual navigation of reformulations, we generate a sample image (``thumbnail'') for each reformulation and display both in a list of suggested reformulations to the user.
The user can choose a suggestion and modify the suggestion before using it to generate images.
This interface is designed to have a comparable interaction level and similar features to \agonistic{} for fair comparison.

\noindent
\textbf{\agonisticbox}~$\cdot$~
The \agonistic{} interface aims to provide resources for users to engage in political controversies over visual representations of their prompt.
Drawing from \S\ref{agonistic-democracy} and \S\ref{hai-reflection}, we take inspiration from \citet{angelietal21agonisticgames} to derive the following tenets of agonistic design for image generation tools: (\textbf{T1}): confront users with unsettling content that challenges user assumptions; (\textbf{T2}): encourage engagement with relevant ongoing political controversies; (\textbf{T3}): respect users as adversaries by allowing them to navigate conflicts on their own terms.

\begin{figure}
    \includegraphics{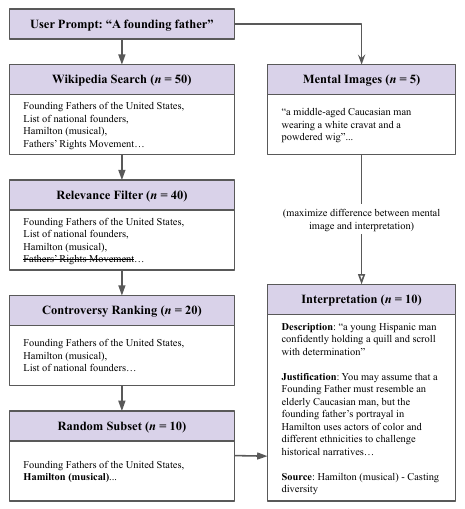}
    \caption{Sample agonistic interface generation workflow.}
    \label{fig:agonistic-workflow-diagram}
\end{figure}

We implement \agonistic{} with a multi-step retrieval-augmented workflow that researches controversies about the user prompt from Wikipedia and presents them to the user (Fig. \ref{fig:agonistic-workflow-diagram}).
We identify the main subject of the prompt with GPT-3.5, perform a Wikipedia search for 50 pages related to the main subject, and use GPT-4o to filter for the 40 pages most relevant to the user prompt by title.
Next, we compute a controversy score for each page by dividing the number of edits by the number of unique editors, following findings from \citet{kitturetal2007conflictwikipedia} that controversy is positively correlated with number of edits and negatively correlated with number of unique editors.
We choose this particular controversy metric for its efficiency and strong qualitative performance in preliminary tests.
Finally, we rank the pages by controversy and select a random subset of 10 from the top 20 most controversial pages.
We select Wikipedia as a research source following because its public editing process lends itself well to capturing the types of public controversies characteristic of agonistic pluralism, as suggested by as \citet{crawford2014agonistic} (\textbf{T2}).

Using this set of 10 pages, we generate interpretations of the user's prompt.
We first instruct GPT-4o to generate 5 mental images an ``average person'' might have of the main subject.
For each page, we instruct GPT-4o to generate an interpretation with 4 fields: 1) a \textit{section summary} explaining the main points of the referenced page section; 2) a \textit{visual description} of the prompt based on information from the section; 3) a \textit{source} with the referenced page and section; 4) a \textit{justification} explaining how the section content justifies the description.
We instruct GPT-4o to generate descriptions the challenge the provided mental images to increase the likelihood that users encounter discomforting suggestions (\textbf{T1}).
The section summary is not shown to the user but is generated first to reduce model hallucinations.
Finally, like in \reformulative, we generate a ``thumbnail'' for each interpretation.
Users are shown a list of interpretations with descriptions, thumbnails, and sources.
When users click on an interpretation, the card expands to show the justification.
After a 3-second wait period to encourage users to read the justification, the user can ``accept'' the interpretation and generate images, as well as edit the description text and re-generate (\textbf{T3}).

\section{Experimental Design}
\label{experiment}

To evaluate how each interface affects user reflection, we conducted a within-subjects comparative lab study and
asked
\textbf{(RQ1):} \textit{How} each interface induces reflection using a variety of measures for user reflection; and
\textbf{(RQ2):} \textit{Why} each interface induces reflection by examining how other  variables could explain the reflection observations in RQ1.

\subsection{User Task and Interview Structure}
\label{task-structure}

Each lab study session was conducted 1-on-1, lasted around one hour, and consisted of the \textbf{setup} and \textbf{user task}.

\textbf{Setup ($\sim$15 mins).}
The interviewer received consent to record the interview and use anonymized data.
Each participant was assigned to a random ordering of interfaces, with \baseline{} fixed as the first interface, to account for learning effects between non-\baseline{} interfaces.
Then, to familiarize the participants with the interfaces, participants were guided through using each interface in their assigned order with the prompt ``a person.''
Afterwards, participants selected a subject (prompt) for their task; they were told to choose one of interest to them and encouraged to choose one in a randomly pre-assigned category.
Subjects all featured people and were divided into three categories: identity/demographics, history, and politics.
This choice of categories allowed for potential reflection on issues of visual representation.
Participants were provided with examples in each category for inspiration but also allowed come up with their own subjects. 
See \S\ref{participant-background} for participants' chosen prompts and \S\ref{topic-list} for example prompts.

\textbf{User Task ($\sim$45 mins).}
After setup, participants were informed of their task: to create a collage of ten images which represented all relevant aspects of the subject.
This task forced participants to make nuanced choices about what to include and exclude in visual representation of their chosen subject.
Participants constructed an initial collage of ten images with \baseline.
Then, they improved their collage by interacting with the other interfaces (\diverse, \reformulative, \agonistic) in a preassigned order. 
If participants produced an image they wanted to add to their collage, they picked one of their existing images to replace.
The task design of accumulating collage improvement gave permanence to users' decisions about visual representation, provided a holistic picture of how users interacted with interfaces across different stages of exploration, and was a more natural and engaging task for participants compared with other experiment designs tested.
Participants were instructed to ``think out loud,'' commenting on their choices to include or exclude images in their collage.
Fig.~\ref{fig:example-collage-progression} displays an example collage progression from the study.

\subsection{Recruiting Procedure and Study Chronology}
\label{recruiting-procedure}

After receiving IRB approval from our university, we sent recruiting materials to undergraduate and graduate students in university departments and student groups via online communication channels.
We aimed to represent a variety of backgrounds, experiences, and interests in our participant pool.
We piloted an initial version of our study with 11 individuals and revised the task to its current form (as described in \S\ref{task-structure}) in response to time constraints and perceived confusion about the task.
Then, the two co-first authors individually conducted interviews with 29 individuals over Zoom.
Participants were compensated $\$20$ per hour.
More information about participants is available in \S\ref{participant-background}.

\subsection{Measures}
\label{measures}

Given the definition of reflection discussed in \S\ref{hai-reflection}, we measured reflection by looking quantitative and qualitative measures of how much individuals questioned or changed prior assumptions after using our interfaces. 
While it is more common to use purely qualitative metrics when measuring reflection \cite{bentvelzenetal2022revisitingreflection}, we believed a mixed-methods approach offered unique benefits for our study because quantitative metrics would allow us to more richly compare reflection across interfaces.
Because of our small sample size, the quantitative methods were selected for the purpose of making \textit{broad observations} about induced reflection rather than precise claims.
We drew from common metrics in mixed methods studies of reflection such as interviews, questionnaires, and self-reports \cite{bentvelzenetal2022revisitingreflection}.

\textbf{Survey.}
After participants interacted with each interface, they provided responses on a 5-point Likert-style scale to measure perceived rethinking (how much their mental image changed), satisfaction, appropriateness, and control (see \S\ref{survey-questions}). 
For the \reformulative{} and \agonistic{} interfaces, we also collected responses about how interesting the suggestions/interpretations were.
Quantitatively measuring self-reported rethinking helped capture reflection not have been recorded by other artifacts like collages (for instance, when an image caused reflection but was not added). 
We measured non-reflection variables (e.g., control) to explain why interfaces might have different reflection.

\textbf{Interview Coding.}
After conducting interviews, the two co-first authors coded interview recordings to systematically extract further information about user experience. 
They independently coded a small subset of interviews before coming together to build a coding ontology.
We divided codes into \textit{intents} (how user intents changed while adding an image, roughly correlated to different degrees of reflection) and \textit{values} (reasons users give for adding an image).
Values are not mutually exclusive, because a user can invoke multiple values when adding an image.
Interviews were coded on the level of individual images added: for each added image, the authors recorded which intent and value codes were invoked by the user.
The two first co-authors then independently coded a subset of three interviews to compute inter-rater reliability (IRR), revising the coding ontology and independently re-coding interviews as needed to resolve disagreements. 
After coding all interviews, the two co-first authors further reviewed all images marked with non-\direct{} intent together to establish higher agreement on intent codes.
The final IRR was 0.67 based on a weighted average of Cohen's Kappa scores across value codes. 
See \S\ref{adding-images} and \S\ref{why-add} for a taxonomy of intent and value codes, and \S\ref{irr-methodology} for the IRR methodology.
\begin{table*}[t]
\centering
\begin{tabularx}{\textwidth}{|l|X|}
\hline
\multirow{2}{*}{\raisebox{-1.3\height} {\rotatebox{90}{\baselinebox}}}
    & \small After observing that all the images for ``the constitutional convention'' featured the delegates sitting, the participant specified that the delegates should be standing. They reflected: ``\textit{when I kept generating those prompts and they were sitting [...] I don't think I would have specified whether they were sitting or standing before. Since I thought about it, I had to... play around with the prompt to get them to stand}''. They further remarked: ``\textit{Interacting with this interface [...] made me pay more detailed attention to what I perceive would have actually happened at that historical event.}'' (\pref{24}) \\ \cline{2-2}
    & \small After generating images for the subject ``a queer community'' which seemed to ``\textit{rely on heavy stereotypes [...] like pride parades}'', the participant wanted to generate ``\textit{everyday examples of queer people}''. After producing a few images, they remarked ``\textit{now you lose the queer part}.'' They reflected: ``\textit{Well now you really got me questioning myself [...] what is a queer community? Do they all have to be wearing pride flags? How do you tell?}'' (\pref{4}) \\ \hline
\multirow{2}{*}{\raisebox{-1.7\height} {\rotatebox{90}{\diversebox}}}
    & \small The participant was attempting to generate an image of ``A Syrian refugee working in Germany'' when \diverse{} generated an image of a woman in a healthcare setting. The participant paused and remarked, ``\textit{can you train to be a doctor [in] 12 years?}'' They later explained that they were unsure whether a Syrian refugee would have been able to study long enough to become a licensed medical professional since leaving Syria, but reasoned that ``\textit{the situation has gone on long enough [that] they have studied and integrated in the societies they're in, where [they're] able to actually finish something in the medical sciences.}'' (\pref{19}) \\ \cline{2-2}
    & \small The participant had previously generated all images of White people for the prompt ``a Jewish person'' using \baseline. After \diverse{} produced images of Black people, the participant felt that the images were not appropriate but became confused upon further reflection: ``\textit{Okay, I'm focused on the Black [...] person here because, this is because my lack of knowledge [...] I don't know what makes a person Jewish other than the religious element [but...] to be Jewish, I think you have to be born in a family [...] I could be wrong.}'' (\pref{29}) \\ \hline
\multirow{2}{*}{\raisebox{-0.95\height} {\rotatebox{90}{\reformulativebox}}}
    & \small With the prompt ``A gun owner,'' \reformulative{} generated the suggestion, ``An elderly man polishing his grandfather's antique shotgun.'' Upon reading the suggestion, the participant reacted, ``\textit{I didn't even think about that kind [of] collector gun owner,}'' before using the suggestion to add an image to their collage, later reflecting that the interface ``\textit{reminded [them] of things [they] forgot about}'' and ``\textit{sparked a new mental image}''. (\pref{10}) \\ \cline{2-2}
    & \small Given the prompt ``A gun control advocate,'' \reformulative{} generated a suggestion of ``A gun control advocate engaging in a community discussion.'' The participant reacted, ``\textit{I like that it's suggested positioning the advocate in a community space, where all of the other images [from previous interfaces] were very portrait-style.}'' They used the suggestion to generate and add an image to their collage, later sharing that they liked how \reformulative{} helped them ``\textit{think of things that [they] wouldn't think to ask.}'' (\pref{19}) \\ \hline
\multirow{2}{*}{\raisebox{-1.6\height} {\rotatebox{90}{\agonisticbox}}}
    & \small After interacting with \agonistic{} on the subject ``World War II'' which showed many less well-known and more local actualizations of WWII, the participant (whose family was strongly impacted by WWII) revised their design statement from ``\textit{WWII involved mass conflicts that...}'' to ``\textit{WWII involved conflicts that...}''. They reflected: ``\textit{I think it just involves conflicts in general now. It's making me rethink that word [mass]: history tends to focus on real large-scale things, but I find that all of them are important to me personally. I feel like it's actually made me reconnect to the way I think about World War II. And instead of trying to project what I would like, think that other people need to see about World War II.}'' (\pref{5}) \\ \cline{2-2}
    & \small The participant initially described their mental image of the prompt ``the signing of the Declaration of Independence'' as ``\textit{A group of [...] white men like Thomas Jefferson standing around a table signing a document.}'' When \agonistic{} generated images of non-White people, drawing from sources like ``Haitian Declaration of Independence,'' the participant observed, ''\textit{when [...] I put  Declaration of Independence, I assume a mental image of the US Declaration of Independence, but so many nations have Declarations of Independence.}'' They continued, ``\textit{It makes me realize [...] this assumption of bias... Oh my gosh, I was so narrow-minded.}'' (\pref{28}) \\ \hline
\end{tabularx}
\caption{Two examples of reflection observed during lab study for each interface.}
\label{tab:qualitative-examples}
\end{table*}

\section{\textit{How} do interfaces induce reflection?} 
\label{how-reflection}

In this section, we aim to paint a holistic picture of \textit{how} reflection occurred under different interfaces by reporting results that directly measure some aspect of reflection collected in the study, including users' self-reported change in mental image (\S\ref{change-mental-image}) and reflection on a per-image level (\S\ref{adding-images}).

\begin{figure}[t]
    \centering
    \includegraphics[height=4.4cm]{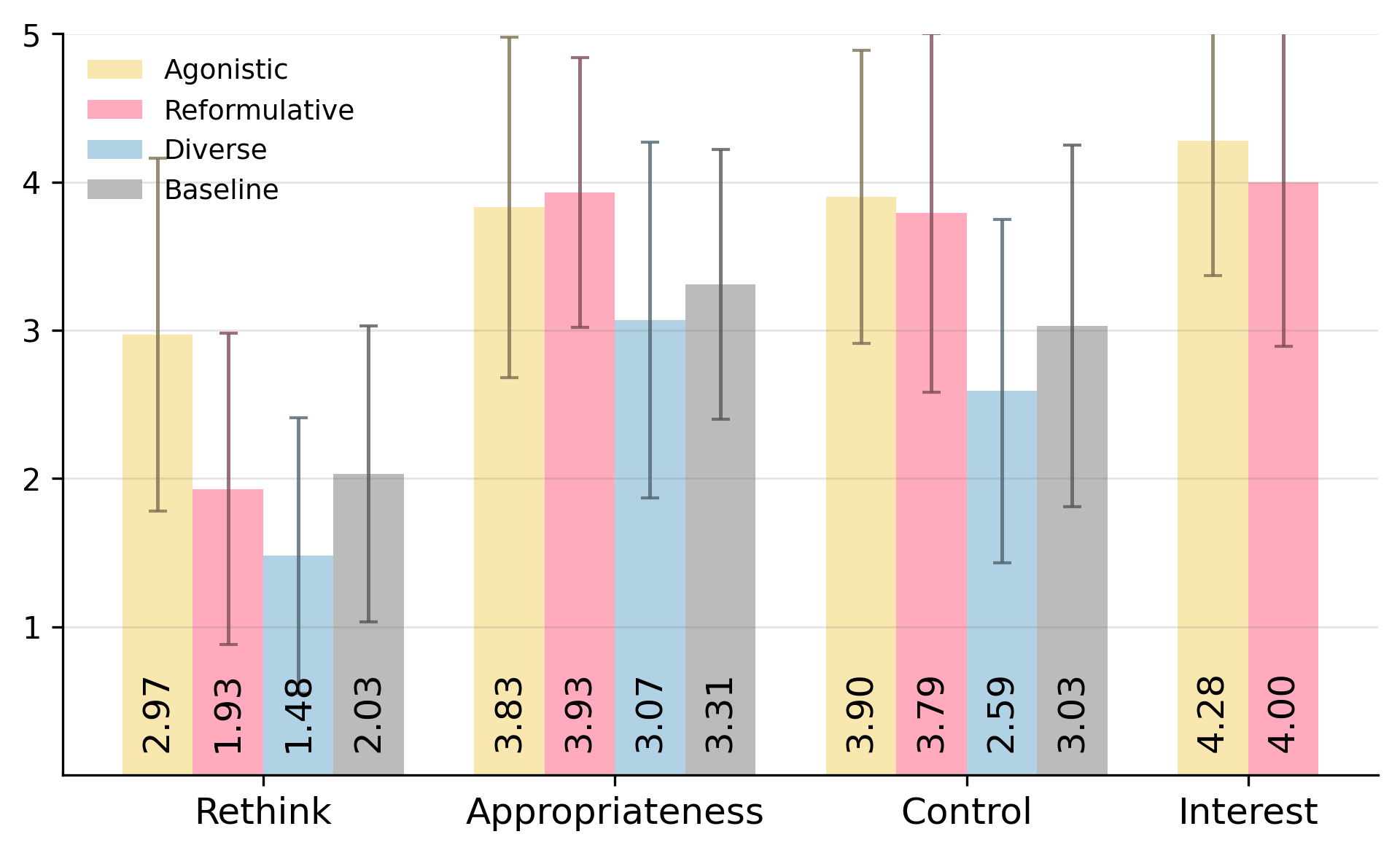}
    \caption{Mean responses on a 5-point scale for Rethink, Appropriateness, Control, and Interest (\S\ref{change-mental-image}, \ref{interface-properties}; Table~\ref{tab:explicit_values_with_std_properties}).}
    \label{fig:properties-visual}
\end{figure}
\begin{figure}[t]
    \centering
    \includegraphics[height=4.3cm]{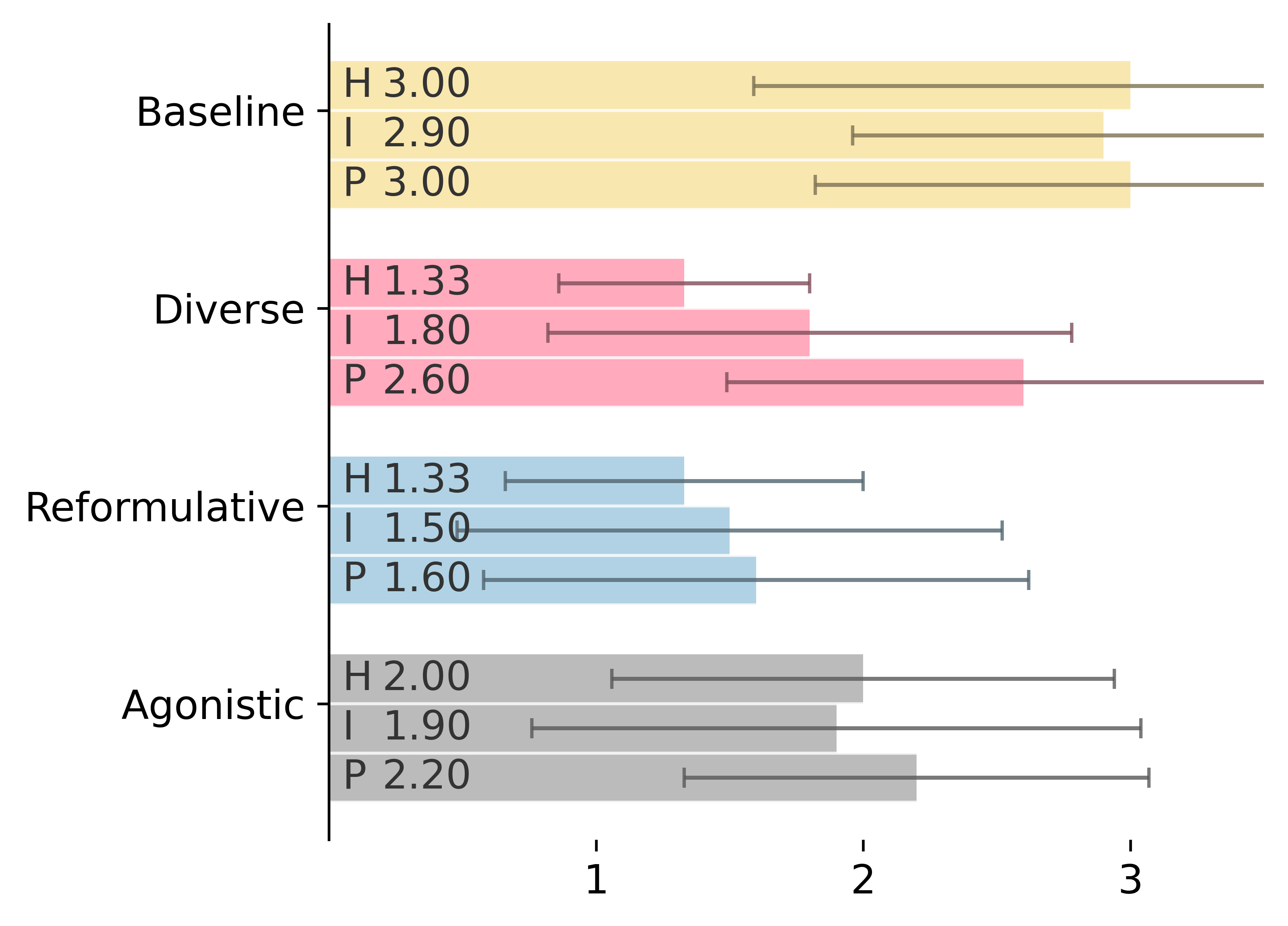}
    \caption{Breakdown of reported rethinking by prompt category: History (`H'), Politics (`P'), Identity (`I') (\S\ref{change-mental-image}; Table \ref{tab:prompt-categories}).}
    \label{fig:prompt-categories}
\end{figure}

\subsection{Change in Mental Picture}
\label{change-mental-image}
Participants self-reported change in mental picture after using each interface by rating agreement with the statement ``Interacting with the interface made me rethink the mental picture of the subject I had right before using this interface'' on a 5-point Likert-style scale.
We interpret higher reported scores as indicative of more reflection.

\textbf{\agonistic{} induces the most reflection as measured by self-reported rethinking (Fig.~\ref{fig:properties-visual}).} 
Participants report greater rethinking under \agonistic{} than \reformulative{} ($p \ll 0.01$), and under \reformulative{} than \diverse{} ($p = 0.05$).
Neither \reformulative{} nor \diverse{} surpass \baseline{}  (likely because participants interacted with \baseline{} first), but \agonistic{} does ($p \ll 0.01$).
The `hierarchy' of interfaces in terms of self-reported reflection is thus $\left[ \agonistic> \{ \baseline, \reformulative\} > \diverse\right]$.

\textbf{Rethinking under \agonistic{} is robust to prompt category (Fig.~\ref{fig:prompt-categories}).} 
Though \diverse{} and \reformulative{} see significant drops in rethinking for historical prompts compared to political prompts ($-0.27$ and $-1.27$, respectively) the same drop is not observed for \agonistic{} ($-0.00$).
For historical topics, rethinking is low for \diverse{} (1.33) and \reformulative{} (1.33), but much higher for \agonistic{} (3.00).
For political topics, rethinking is high for both \agonistic{} (3.00) and \reformulative{} (2.60).
While all prompt categories have roughly similar rethinking under \agonistic, political subjects lend themselves to more rethinking overall under \diverse{} and \reformulative{} than historical subjects.

\subsection{Reflection Induced by Adding Images}
\label{adding-images}

\begin{figure}[b]
    \centering
    \includegraphics[height=4.4cm]{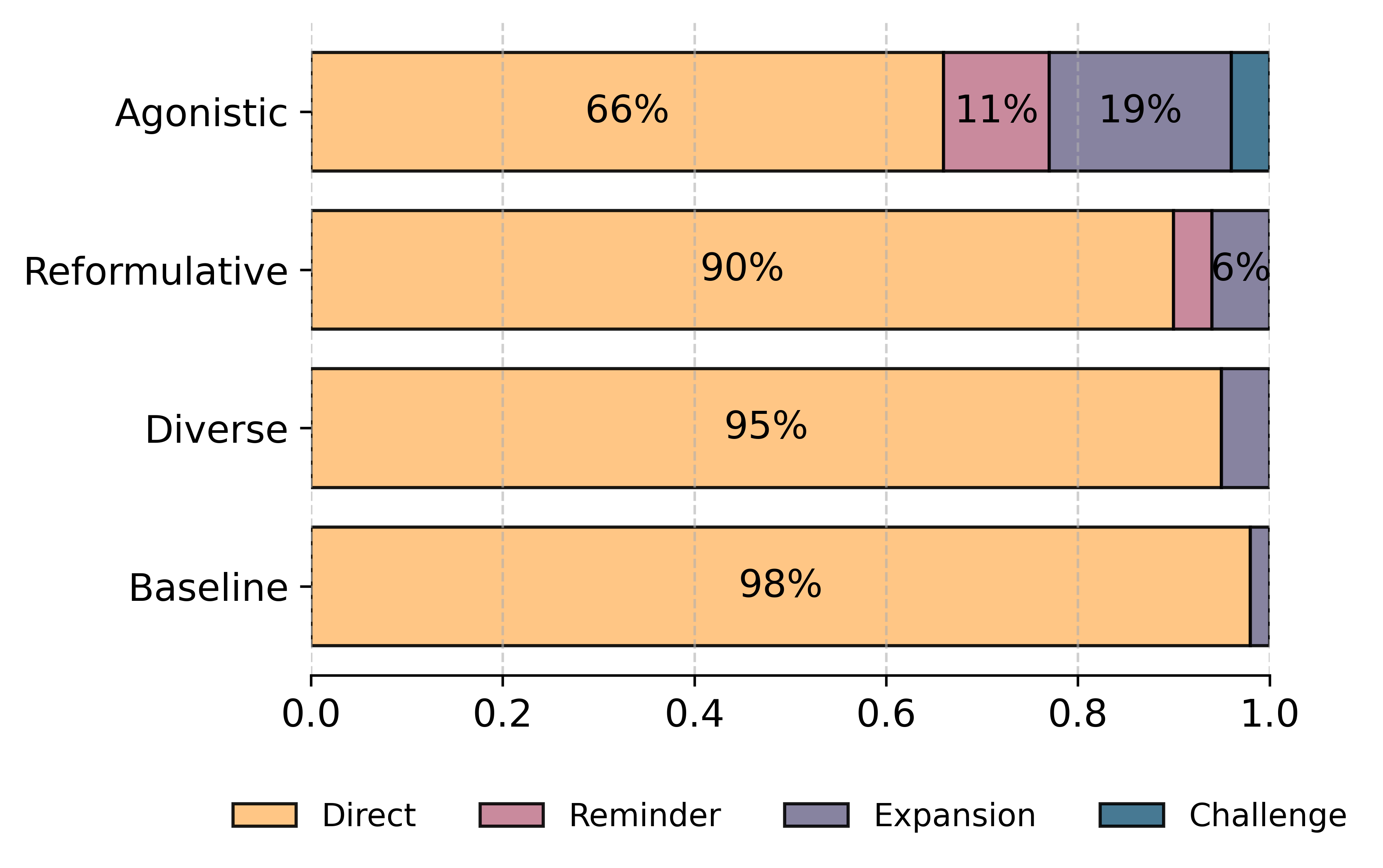}
    \caption{Breakdown of intent codes per interface (\S\ref{adding-images}; Table~\ref{tab:intent-distribution-table}).}
    \label{fig:intents_distributed}
\end{figure}

Next, we analyze the distribution of ``intent'' codes across interfaces. 
As explained in \S\ref{measures}, ``intents'' code for how user intents changed while adding an image to their collage.
We define the following intent codes: 1) \direct: the user already intended to generate the image; 2) \reminder: the user is reminded of an aspect the prompt they already accepted but forgot; 3) \expansion: the user accepts an image as a different valid interpretation of the prompt than their original intent; 4) \challenge: the user realizes an error or otherwise experiences a significant change in their original intent. \direct{} intent represents little to no reflection, whereas \reminder, \expansion, and \challenge{} represent increasing degrees of reflection.

\textbf{\agonistic{} induces the most reflection as measured by quantitative interview analysis (Fig.~\ref{fig:intents_distributed}).}
Participants replace the most images on average with \agonistic{} (3.1), followed by \reformulative{} (2.9) and \diverse{} (1.7).
We observe \direct{} intent most frequently with \baseline{} ($98\%$), followed by \diverse{} ($95\%$), \reformulative{} ($90\%$), and a sharp drop in \agonistic{} ($67\%$). \agonistic{} is associated with the highest proportion of \reminder{} and \expansion{} codes across all interfaces ($11\%$ and $19\%$ respectively, compared to the next-highest $4\%$ and $6\%$ for \reformulative), and is the only interface to \challenge{} intent ($4\%$). 
Table~\ref{tab:qualitative-examples} shows two examples of \challenge{} for \agonistic. 
A minor difference with \S\ref{change-mental-image} is that \diverse{} induces more frequent reflection than \baseline, which might be explained by the fact that our coding schema excluded forms of reflection that did not explicitly correlate to an added image.
The `hierarchy' of interfaces in terms of coded reflection is thus $\left[ \agonistic > \reformulative > { \baseline, \diverse}\right]$.
\section{\textit{Why} do interfaces induce reflection?}
\label{why-reflection}

To explain the observed differences in reflection under different interfaces as \S\ref{how-reflection}, we examine the relationship between reflection and other variables: perceived interface-level properties like appropriateness and control (\S\ref{interface-properties}), values for adding images (\S\ref{why-add}), values for rejecting images (\S\ref{why-reject}), and qualitative themes (\S\ref{qualitative-themes}).

\subsection{Perceived Interface-Level Properties}
\label{interface-properties}

\textbf{\reformulative{} and \agonistic{}  surpass \baseline{} and \diverse{}  in appropriateness and control ($p \le 0.05$ for all) (Fig.~\ref{fig:properties-visual}).}
There is generally a weak but significant correlation between rethinking and both appropriateness ($r = 0.22$, $p = 0.02$) and control ($r = 0.27$, $p = 0.003$).
\agonistic{} is ranked above \diverse{} by $81\%$ of participants for appropriateness and $86\%$ for control, while \reformulative{} is ranked above \diverse{} by $85\%$ of participants for appropriateness and $92\%$ for control among participants.
Although the difference between \diverse{}  and \baseline{}  is not significant for appropriateness, it is for control: participants report lower control with \diverse{} than with \baseline{}  ($p \ll 0.01$), suggesting that rewriting user prompts to increase diversity results in a noticeable feeling of decreased control.
The difference between \agonistic{} and \reformulative{} is not significant for control or appropriateness ratings ($p \gg 0.05$), but participants find \agonistic's interpretations more interesting than \reformulative's suggestions ($p \ll 0.05$).

\subsection{Why Users Add Images}
\label{why-add}

We also explore whether ``value'' codes may serve as a relevant explanatory variable for types of reflection (``intent'' codes). As described in \S\ref{measures}, ``values'' code for \textit{why} participants add images, or more specifically the reasons that participants invoke when adding images. The values in our coding ontology are: 1) \realism: the image represents how the user thinks the world actually is; 2) \familiarity: the image fits the user's assumptions, regardless of whether the user thinks they are representative of how the world actually is; 3) \diversity: the image portrays an underrepresented aspect of the prompt that the user believes is normatively important to include; 4) \aesthetics: the user likes how the image looks.

\begin{figure}[t]
    \centering
    \includegraphics[height=4.4cm]{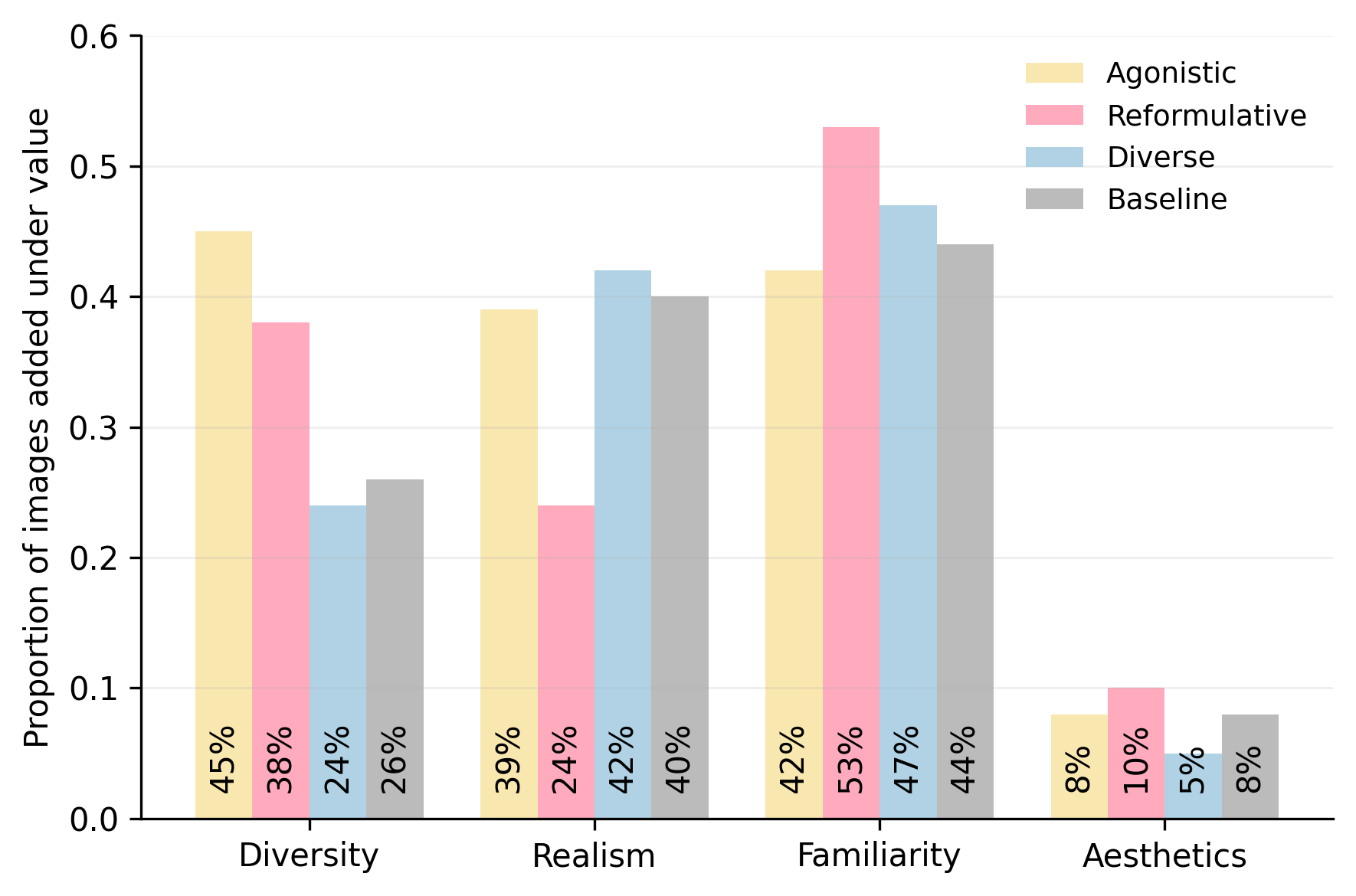}
    \caption{Distribution of value codes per interface (\S\ref{why-add}; Table~\ref{tab:values-interfaces}).}
    \label{fig:values-interfaces}
\end{figure}

\textbf{\agonistic{} is the only interface for which \diversity{} is invoked more frequently than \familiarity{} (Fig.~\ref{fig:values-interfaces}).} In all interfaces \textit{except} \agonistic, \familiarity{} is invoked most frequently, followed by \realism, \diversity, and \aesthetics. For \agonistic, participants invoke \diversity{} most frequently, followed by \realism, \familiarity, and \aesthetics. Compared to \baseline, participants invoke \diversity{} \textit{less} frequently for \diverse{} ($-2\%$), but more frequently for \reformulative{} ($+12\%$) and \agonistic{} ($+19\%$), suggesting that \diverse{} is not an effective way to portray human diversity.

\begin{figure}[t]
    \centering
    \includegraphics[height=4.4cm]{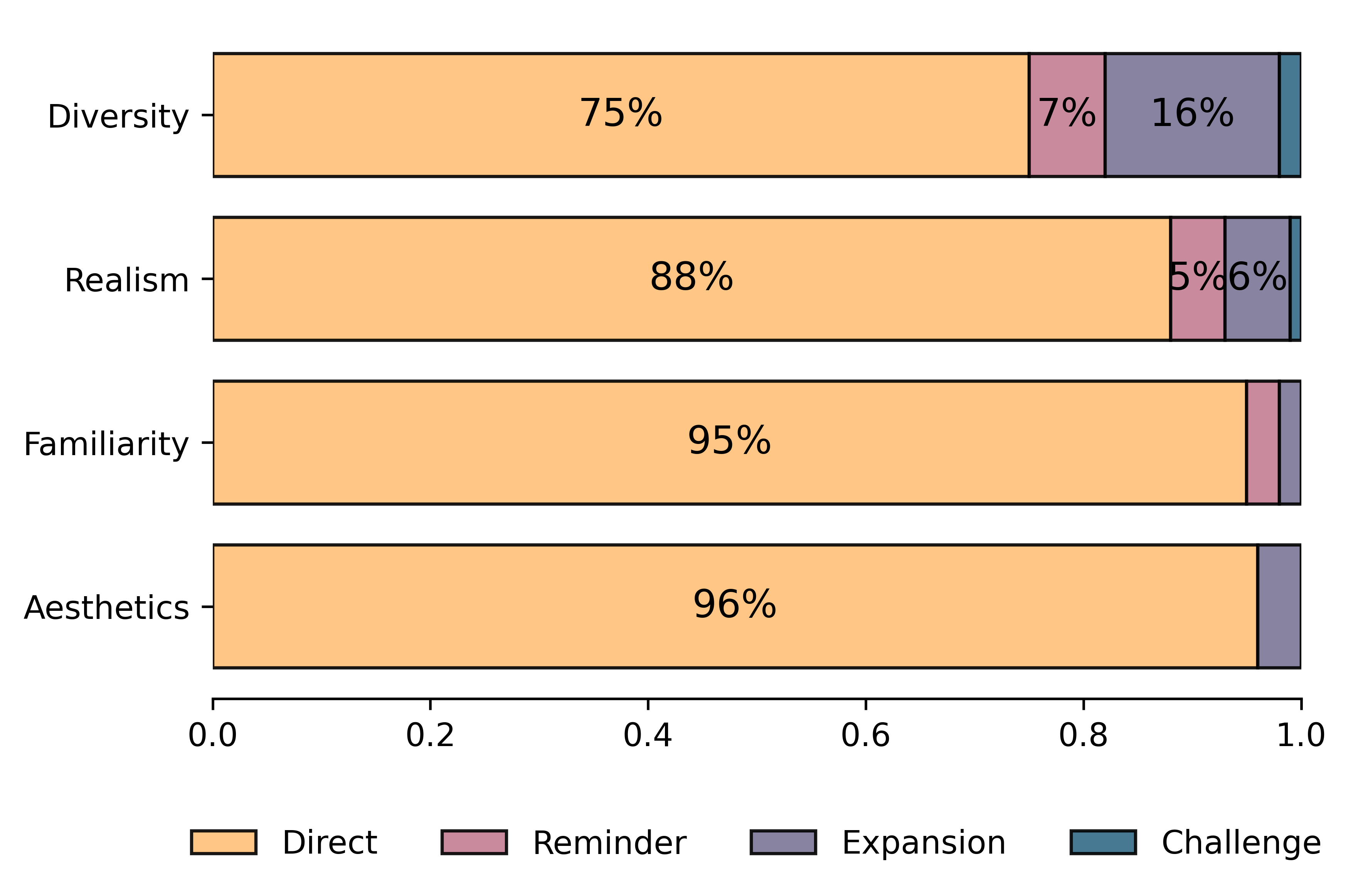}
    \caption{Breakdown of co-occurrence between intent codes (x-axis) and value codes (y-axis) (\S\ref{why-add}; Table~\ref{tab:values-intents-rel}).}
    \label{fig:intents-values}
\end{figure}

\textbf{\diversity{} is associated with the highest rate of reflection (Fig. \ref{fig:intents-values}).}
For all values, the majority of images are added with \direct{} intent. \aesthetics{} is associated with the highest proportion of \direct{} intent ($96\%$), followed by \familiarity{} ($95\%$), \realism{} ($89\%$), and \diversity{} ($75\%$). \diversity{} is associated with the highest degrees of reflection, with $7\%$ of images added with \reminder, $16\%$ of images added with \expansion, and $2\%$ of images added with \challenge{} of intent. Since \agonistic{} also has the highest proportion of \diversity{} values, these findings suggest that \diversity{} serves as a strong explanatory variable for reflection.

\subsection{Why Users Reject Images}
\label{why-reject}

We carry out a qualitative analysis of image rejections in lieu of a quantitative analysis because the unit of rejection is not as well-defined as added images.

\textbf{Participants reject images due to factual inaccuracies, especially when interacting with \diverse.}
Using \diverse, \pref{7} said that racially diverse pictures of ``The Chinese Communist Revolution'' ``\textit{looked like modern people were slapped into the wrong time period,}'' and \pref{28} experienced ``\textit{pushback}'' against racially diverse pictures of ``The Signing of the Declaration of the Independence,'' similar to public pushback in the Gemini controversy.
Participants similarly cited lack of factuality as a reason for rejecting images in other interfaces, though to a lesser extent.

\textbf{Participants also reject images because of an excessive adherence to factuality.} 
With the prompt ``A Syrian refugee,'' \pref{26} noted, ``\textit{[\agonistic] presented more of what I know to be true, [...] the harsh reality of it. [...] The other interfaces [...] were more instrumental in helping me expand beyond that}.'' 
\pref{7} preferred \reformulative{} over \agonistic{} because \reformulative{} did a better job `\textit{`highlighting all different aspects of [the prompt], instead of just the realism images.}'' These comments suggest that participants value some level of diversity in generated images beyond what they consider to be strictly factual.

\textbf{Participants reject images due to lack of familiarity, even where they acknowledge that an image may be factual.} 
When using \agonistic, some participants believed interpretations to be factual (being based on Wikipedia) but still rejected them because they are unfamiliar.
For example, with the prompt ``An Arab person,'' \pref{25} reacted to an image of a Sudanese Arab person from \agonistic: ``\textit{Sudanese Arab [...] that's an identity that I haven't represented here that I would like to.}'' 
However, they later decided to reject the image: ``\textit{It just does not resonate with me. I guess I'm also not familiar with Sudanese Arabs as much.}''

\subsection{Qualitative Themes}
\label{qualitative-themes}

We present several themes from qualitative analysis that further characterize differences in reflection observed in \S\ref{how-reflection}.

\textbf{Text helps users contextualize, interpret, and decide on images.} 
We find that participants prefer interfaces that help them contextualize and interpret images (\reformulative{} and \agonistic{} over \baseline{} and \diverse), even though ultimately they are only adding the images (and not the accompanying text) to the collage.
Contextualizing content is cognitively stimulating, with \pref{24} noting that the ``\textit{suggestion feature [...] makes your brain think about alternatives}'' and \pref{1} saying that they 
``\textit{wouldn’t have thought of [these images] without the text}.''
Participants report that text is important for mediating user interpretations of images.
\pref{1} said that the generated images ``\textit{adhere to the prompt but you have to figure out the image yourself}.''
\pref{15} said of \reformulative, ``\textit{I like with this interface, it gives me a lot more of an explanation. It [...] tells a story that I might not know just by looking at it [and] could help me kind of change what I'm looking for.}''

\textbf{``Authentic'' diversity is situated in broader political contexts.}
Although both \diverse{} and \agonistic{}  display demographically diverse images to users, participants overwhelmingly prefer the diversity of \agonistic{}  to that of \diverse{} due to a greater perception of \textit{authenticity}. 
Participants perceive \diverse{} as not only factually inaccurate (\S\ref{why-reject}), but also opaque, alienating, and disingenuous.
\pref{17} said they couldn't ``\textit{get a clear idea from anything}'' from \diverse, while \pref{5} reacted to racially and gender diverse World War II soldiers from \diverse{} in the following way: ``\textit{I find them not just [...] historically inaccurate, I think it gives narratives that some people will not dig deeper into, and [...] can lead to a lot of misconceptions down the line of colorblind racism.}''

In contrast, participants report that the perceived authenticity of \agonistic{} engages them in the political context of image generation.
\pref{10} liked that \agonistic{} was ``\textit{more grounded in reality [...] you're like, `yeah, real people exist.'}''
\pref{28} shared, ``\textit{[\agonistic] changed my perception of what the prompt itself was [...] but [\diverse] feels like it's [...] trying to change our perception of the event itself that I was already thinking of [...\agonistic] tries to bring in diverse perspectives and increase representation of other groups [...] in a way that felt more \textbf{authentic} and can change people's assumptions in a healthier way.}'' 

\textbf{\agonistic{} treads the line between appropriate and inappropriate.}
By design, \agonistic{} often presents users with interpretations that are \textit{initially} perceived as discomforting, inappropriate, or even offensive but may become appropriate after reflection.
However, \agonistic's interpretations may sometimes remain inappropriate even after reflection.
\pref{4} said, ``\textit{[\agonistic] was giving me the most appropriate, but it was also giving me the most inappropriate [outputs] at the same time [...It] would sometimes hit a home run, sometimes wildly off track.}''
\pref{16} said that \agonistic{} was ``\textit{good good and bad bad [...] I like the specificity [of \agonistic]. I just think the specificity was kind of whacked out.}''
These comments reflect the dual nature of agonistic design which by nature will tread a thin line between appropriate and inappropriate content.
\section{Discussion}
\label{discussion}

\subsection{The Inefficacy of Inauthentic Diversity}
\label{inauthentic-diversity}

Since \diverse{} was loosely modeled after the reported behavior of the Gemini case, it provides an interesting point of analysis on the question of diversity in image generation interfaces.
On one hand, some participants prefer \diverse{} to \baseline{} because they feel that the demographic diversity of \diverse{} is either more appropriate for representing the subject or more familiar to their experience. 
In response to images of a diverse crowd worshiping Jesus, \pref{6} said,``\textit{It's nice having [...] diverse people, 'cause I think that's just how it is.}''
On the other hand, as discussed in \S\ref{why-reject}, many participants echo concerns about conflicts between diversity and factuality---diversity's ``factuality tax''~\cite{wang-etal-2023-t2iat}---that also surfaced in the online reaction to the Gemini case~\cite{milmoandkern2024gemini, robertson2024google}.

Nonetheless, participant reactions to \agonistic{} also challenge conventional understandings of the tradeoffs between factuality and diversity. 
For example, when creating a collage for ``the women's suffrage movement,'' \pref{18} was hesitant to include images of non-White women because many American suffragette leaders espoused racist views, but changed their mind after engaging with an interpretation from \agonistic{} that highlighted the role of Black women in leading their own suffragette organizations.
Here, adding the image did not obscure the facts of historical racism but highlighted other relevant aspects in the complex history of the suffragette movement.
Thus although the ``factuality tax'' is perceived as a necessary cost of diversity, it is not clear that the two are always opposed to one another.
Given that images are situated in sociopolitical context, developers might turn attention from a narrow focus on generating only the right kinds of \textit{images} to also generating the right kinds of \textit{context}---something that study participants found valuable (\S\ref{qualitative-themes})---which can give rise to new ways of seeing and understanding the prompt.

These and other reactions point to \textit{authenticity} as a crucial yet under-discussed part of the complex interplay between diversity and factuality.
Taking a definition that ``\textit{[t]o say that something is authentic is to say that it is what it professes to be [...] in origin or authorship}''~\cite{sep-authenticity}, \diverse{} seems not only unfactual but also insufficiently aligned with the underlying broader goals of diversity: that is, \textit{representing and respecting underrepresented populations in ways that matter to them}.
In \S\ref{qualitative-themes}, we documented how situating diversity in a political context encourages user perception of authenticity, as further exemplified by \pref{18}'s reflection on the suffragette movement above.
Moreover, our findings that participants invoke \diversity{} at higher rates when adding images with \agonistic{} than \diverse{} (\S\ref{why-add}) suggest that agonistic pluralism offers a superior framework for actualizing authentic diversity by engaging users with this political context rather than imposing top-down interpretations upon them.
These findings further support the idea that the ``factuality tax'' is an overly restrictive framework for conceptualizing \textit{authentic} diversity.
Authentic diversity should be understood as neither diversity for the sake of diversity nor diversity rigidly constrained by factuality, but as the difference of interpretations that emerges out of real human political conflict.

\subsection{Engaging with Users Experientially}
Our findings highlight another component of user engagement missing from the factuality-diversity dichotomy: the role of \textit{familiarity}.
As discussed in \S\ref{why-add} and \S\ref{why-reject}, participants often accept and reject images because they are un/familiar rather than because they are un/factual or un/diverse.
To steer users towards desired ideals like critical reflection, interfaces should treat them not solely as judges for factuality and diversity, but as people who engage with images on the basis of familiarity and experience.
The design of \reformulative, for example, leans into giving users aestheticized and familiar experiences, possibly at the expense of reflection:
``\textit{if there was a change that I wanted to implement in these pictures, but I just didn’t have the words for it, [\reformulative] gave the options. \textbf{I didn’t have to [...] think about what I was looking for}. I could just see examples of what I was looking for}'' (\pref{19}).
Meanwhile, \agonistic{} focuses not only on producing factual and diverse outputs via Wikipedia-grounded interpretations, but also explicitly engages and challenges users' experiential sense of familiarity.
This reflects yet another core principle of agonistic pluralism, which emphasizes the ``\textit{crucial role played by passions and affects in securing allegiance to democratic values}''~\cite{mouffe2000democraticparadox}.

\subsection{The Sponteneity of Reflection}
Despite the differences in reflection highlighted above, participants generally experienced at least \textit{some} level of reflection across all interfaces (as demonstrated by Table~\ref{tab:qualitative-examples}),
indicating that reflection is a highly subjective, personal experience.
Indeed, while \baseline{} was uninspiring for many---``\textit{I typed what I thought and I got what I thought I would get}'' (\pref{21})---it was a highly reflective experience for others---``\textit{I have a lot of other potential images in my mind of it now, and I don't know what is [...] actually true}'' (\pref{28}).
Although reflection often occurs in planned ways (i.e., as a result of design elements we introduced with the intention of inducing reflection), it also often emerges in spontaneous yet personally significant ways as encounters with the Other~\cite{ye2024languagemodelscriticalthinking, halbertandnathan2015criticalreflection, levinas_totality_1969, lacan_ecrits}.

\subsection{Limitations and Future Work}
Wikipedia entries are overtly sites of discourse and negotiation, and prone to political, cultural, and other forms of bias~\cite{hube2017bias, umarova2019partisanship, Greenstein2014DoEO}.
Even though the \agonistic{} generation workflow is designed to output a diverse range of interpretations, this bias may nonetheless filter into the process in a way that hinders agonistic user engagement.
Future work might seek to explore ways to design AI interfaces that allow users to engage more deeply with the discursive processes like Wikipedia edit history by which these online communities produce knowledge.
Similarly, future work on agonistic design might also consider more explicit and iterative representations of discourse (e.g., graph-based), rather than \agonistic's summarizing and linear presentation of discourse/interpretations.

In addition, we recognize that agonistic design may not be suited to all applications.
We have argued that agonistic pluralism is particularly useful at challenging dominant social assumptions, but for this reason \citet{halbertandnathan2015criticalreflection} also caution that critical reflection is less helpful for marginalized populations who already regularly face discomforting experiences in their everyday lives.
Likewise, since our sample population was recruited from students at a single university, our study may be prone to our study may be prone to novelty and participant response biases~\cite{dell2012yoursbetter}.
For example, university students may have already been predisposed to critical reflection from classroom experiences, and students at an urban institution may have already been more receptive to diverse forms of representation.
Since our study favored lower-scale but highly-detailed examinations of user experiences, our results may therefore be limited in generalizability to other populations.
We hope that this paper sparks further explorations of agonistic design in other contexts to better understand its limits and effectiveness at encouraging critical reflection for other populations.
\section{Coda: Recentering Conflict in Diversity}
\label{conclusion}

From image generation interfaces emerges a \textit{political world} in which individuals negotiate over the meanings of images.
We introduced \textit{agonistic pluralism} as a challenge to the ``hegemony of intention'' in current image generation interface design, demonstrating that an interface built on agonistic principles can engage individuals in this political world, making them more reflective.
In her seminal essay ``Situated Knowledges,'' Donna Haraway uses the metaphor of vision to call for the need to ``\textit{attach the objective to our theoretical and political scanners in order to name where we are and are not.}''
``\textit{The moral is simple,}'' Haraway writes: ``\textit{only partial perspective promises objective vision}'' \cite{haraway1988situated}.
By engaging users with the larger discourse they enter into upon prompting an image generation model, our interface aims to activate users' ``theoretical and political scanners'' to see beyond what they intended.
Agonistic interfaces surface a multitude of partial perspectives for a more ``objective'' user experience of image generation---more engaged with the ideas, tensions, and imaginations of real people around them.

Ultimately then, our study highlights the need for \textit{recentering conflict in diversity}---not just comfortable diversity that ``doesn't make a difference,'' but unsettling diversity that challenges our partial perspectives.
Why is it important to care about the demographic diversity of doctors, CEOs, and Founding Fathers?
Not because these cases fall under some objective standard of diversity, but because they are the products of ``\textit{real acts of force}'' that have shaped ongoing \textit{human} conflicts over representation, both in the semiotic and political sense \cite{melamed14diversity}.
Contrary to conventional narratives of diversity as a harmonious ideal, agonistic pluralism recognizes that diversity in its most authentic form stems from genuine and irreconcilable conflicts over power and representation.
If such disagreements are indeed a fact of life as Mouffe suggests, then the radically democratic response for technologists is not to rush to placate user intentions by avoiding sociopolitical controversies, but rather to confront confront controversy head-on by repoening spaces where democratic conflict can take place.

\begin{acks}
This project was partially funded by Sony Faculty Innovation Award.
\end{acks}

\bibliographystyle{ACM-Reference-Format}
\bibliography{main}

\onecolumn
\newpage
\appendix

\section{Full Study Materials and Information}

\subsection{Participant Backgrounds}
\label{participant-background}

Participants came from the following departments:
Computer Science,
Physics,
Math,
Economics,
Informatics,
Comparative History of Ideas,
English,
History,
Business,
American Ethnic Studies,
Law,
Journalism,
Education,
Library and Information Sciences,
Gender Women and Sexuality Studies,
Communications,
Cinema and Media Studies,
and
Political Science.
Out of the 29 participants in the study, 21 were undergraduate students and 8 were graduate students.

\begin{table}[h!]
\centering
\begin{tabular}{|c|l|l|l|}
\hline
\textbf{PID} & \textbf{Selected Prompt}                             & \textbf{Category} & \textbf{Interface Order}              \\ \hline
\pref{1}  & A Korean person                    & Identity   & \diverse, \reformulative, \agonistic \\ \hline
\pref{2}  & A middle-class person              & Identity   & \agonistic, \diverse, \reformulative \\ \hline
\pref{3}  & A designer                         & Identity   & \reformulative, \diverse, \agonistic \\ \hline
\pref{4}  & A community of queer people        & Identity   & \agonistic, \reformulative, \diverse \\ \hline
\pref{5}  & World War II                       & Politics   & \reformulative, \agonistic, \diverse \\ \hline
\pref{6}  & Jesus                              & History    & \diverse, \agonistic, \reformulative \\ \hline
\pref{7}  & The Chinese Communist Revolution   & History    & \diverse, \agonistic, \reformulative \\ \hline
\pref{8}  & An Asian person                    & Identity   & \agonistic, \diverse, \reformulative{} \\ \hline
\pref{9}  & An immigrant                       & Identity   & \agonistic, \reformulative, \diverse \\ \hline
\pref{10} & A gun owner                        & Politics   & \diverse, \reformulative, \agonistic \\ \hline
\pref{11} & The French Revolution              & History    & \agonistic, \reformulative, \diverse \\ \hline
\pref{12} & A fascist                          & Politics   & \reformulative, \diverse, \agonistic \\ \hline
\pref{13} & A socialist                        & Politics   & \diverse, \agonistic, \reformulative \\ \hline
\pref{14} & The Israel-Palestine Conflict      & Politics   & \reformulative, \agonistic, \diverse \\ \hline
\pref{15} & A mass shooter                     & Politics   & \diverse, \reformulative, \agonistic \\ \hline
\pref{16} & A gun control advocate             & Politics   & \reformulative, \diverse, \agonistic \\ \hline
\pref{17} & World War II                       & History    & \agonistic, \diverse, \reformulative \\ \hline
\pref{18} & The women's suffrage movement      & History    & \agonistic, \diverse, \reformulative \\ \hline
\pref{19} & A Kamala supporter                 & Politics   & \reformulative, \diverse, \agonistic \\ \hline
\pref{20} & Lord Krishna                       & History    & \diverse, \reformulative, \agonistic \\ \hline
\pref{21} & An activist                        & Identity   & \agonistic, \reformulative, \diverse \\ \hline
\pref{22} & A Democrat                         & Politics   & \reformulative, \agonistic, \diverse \\ \hline
\pref{23} & A Roman gladiator                  & History    & \agonistic, \diverse, \reformulative \\ \hline
\pref{24} & The Constitutional Convention      & History    & \reformulative, \agonistic, \diverse \\ \hline
\pref{25} & An Arab person                     & Identity   & \reformulative, \diverse, \agonistic \\ \hline
\pref{26} & A Syrian refugee                   & Politics   & \diverse, \reformulative, \agonistic \\ \hline
\pref{27} & A Christian person                 & Identity   & \diverse, \agonistic, \reformulative \\ \hline
\pref{28} & Declaration of Independence Signing& History    & \agonistic, \reformulative, \diverse \\ \hline
\pref{29} & A Jewish person                    & Identity   & \diverse, \agonistic, \reformulative \\ \hline
\end{tabular}
\caption{Participant IDs with their selected prompt, category, and assigned order for non-\baseline{} interfaces.}
\label{tab:prompts}
\end{table}

\newpage

\subsection{Full Topic List}
\label{topic-list}

Participants were provided the following list of example topics/subjects within each of the three categories.

\begin{itemize}[label=$\bullet$, leftmargin=1.5em]

    \item \textbf{Identity \& Demographics}
    \begin{itemize}[label=$\circ$]
        \item \textit{Race:} ``a Black person'', ``a White person'', ``an Asian person''
        \item \textit{Ethnicity:} ``a Hispanic person'', ``a Jewish person'', ``an Arab person''
        \item \textit{Nationality:} ``an American person'', ``a Chinese person'', ``a Brazilian person'', ``a Swedish person'', ``a South African person''
        \item \textit{Class:} ``a working class person'', ``a middle class person'', ``a member of the elite''
        \item \textit{Family history:} ``an immigrant'', ``a refugee'', ``an indigenous person''
        \item \textit{Occupation:} ``a nurse'', ``a student'', ``a businessperson'', ``a CEO'', ``a/the Pope''
        \item \textit{Religion:} ``a Christian person'', ``a Muslim person'', ``a Buddhist person''
        \item \textit{LGBTQ+:} ``a transgender person'', ``two men getting married''
    \end{itemize}

    \item \textbf{History}
    \begin{itemize}[label=$\circ$]
        \item \textit{Historical people:} ``a Founding Father'', ``a Nazi soldier'', ``Jesus'', ``the prophet Muhammad'', ``a Roman gladiator''
        \item \textit{Historical events:} ``World War I'', ``the French Revolution'', ``the Holocaust'', ``the Great Depression'', ``the American Civil War'', ``the Chinese Communist Revolution'', ``the fall of the Berlin Wall'', ``the Trail of Tears'', ``D-Day'', ``The Declaration of Independence Signing'', ``the Nakba'', ``the Napoleonic Wars''
    \end{itemize}

    \item \textbf{Politics}
    \begin{itemize}[label=$\circ$]
        \item \textit{Ideological orientations:} ``a Republican'', ``a Democrat'', ``a rightist'', ``a leftist'', ``a centrist'', ``a Trump supporter'', ``a Harris supporter'', ``a libertarian'', ``a progressive'', ``a nationalist'', ``an anarchist'', ``a socialist'', ``a capitalist'', ``a communist'', ``a fascist'', ``an authoritarian''
        \item \textit{Current conflicts and issues:}
        \begin{itemize}[label=$-$]
            \item \textit{Geopolitical:} ``the Israel-Palestine conflict'', ``the Ukraine-Russia war'', ``a protester in Hong Kong'', ``a Kurdish fighter'', ``a Syrian refugee''
            \item \textit{Gun Rights and Policing:} ``a mass shooter'', ``a gun owner'', ``a gun control advocate'', ``a mass shooting survivor'', ``a police officer in a high-crime area'', ``a police brutality victim''
            \item \textit{Immigration and Borders:} ``a migrant crossing the US border'', ``a border patrol agent''
        \end{itemize}
    \end{itemize}

\end{itemize}

\newpage

\subsection{Full Survey Questions}
\label{survey-questions}

We ask users to provide responses on a 5-point Likert-style scale (1 $=$ ``not at all'', 3 $=$ ``somewhat'', 5 = ``entirely'') after interacting with each interface for each of the following statements:

\begin{itemize}
    \item \textit{Satisfaction}: ``I am satisfied with the collage I created''
    \item \textit{Rethinking}: ``Interacting with the interface made me rethink the mental picture of the subject I had right before using this interface''
    \item \textit{Appropriateness}: ``I thought the content generated from the interface (e.g. images, suggestions) was appropriate to the prompt''
    \item \textit{Control}: ``I feel like the interface gives me detailed control over the image generation process''
    \item  \textit{Interest}: ``I found the content of the suggestions interesting''
\end{itemize}

At the end of the study, participants were also asked to rank each of the interfaces by agreement with the rethinking, appropriateness, and control statements.
This provided another data signal as well as a moment for participants to reflect upon the entire interview and make open-ended comments.
Participants were allowed to give two interfaces the same rank if there were strong ties.

\newpage

\section{Detailed View of Interfaces}
\label{detailed-view-interfaces}

\subsection{Design Iteration for \agonistic}
\label{design-iteration}

\begin{figure}[ht]
    \centering
    \includegraphics[width=0.75\textwidth]{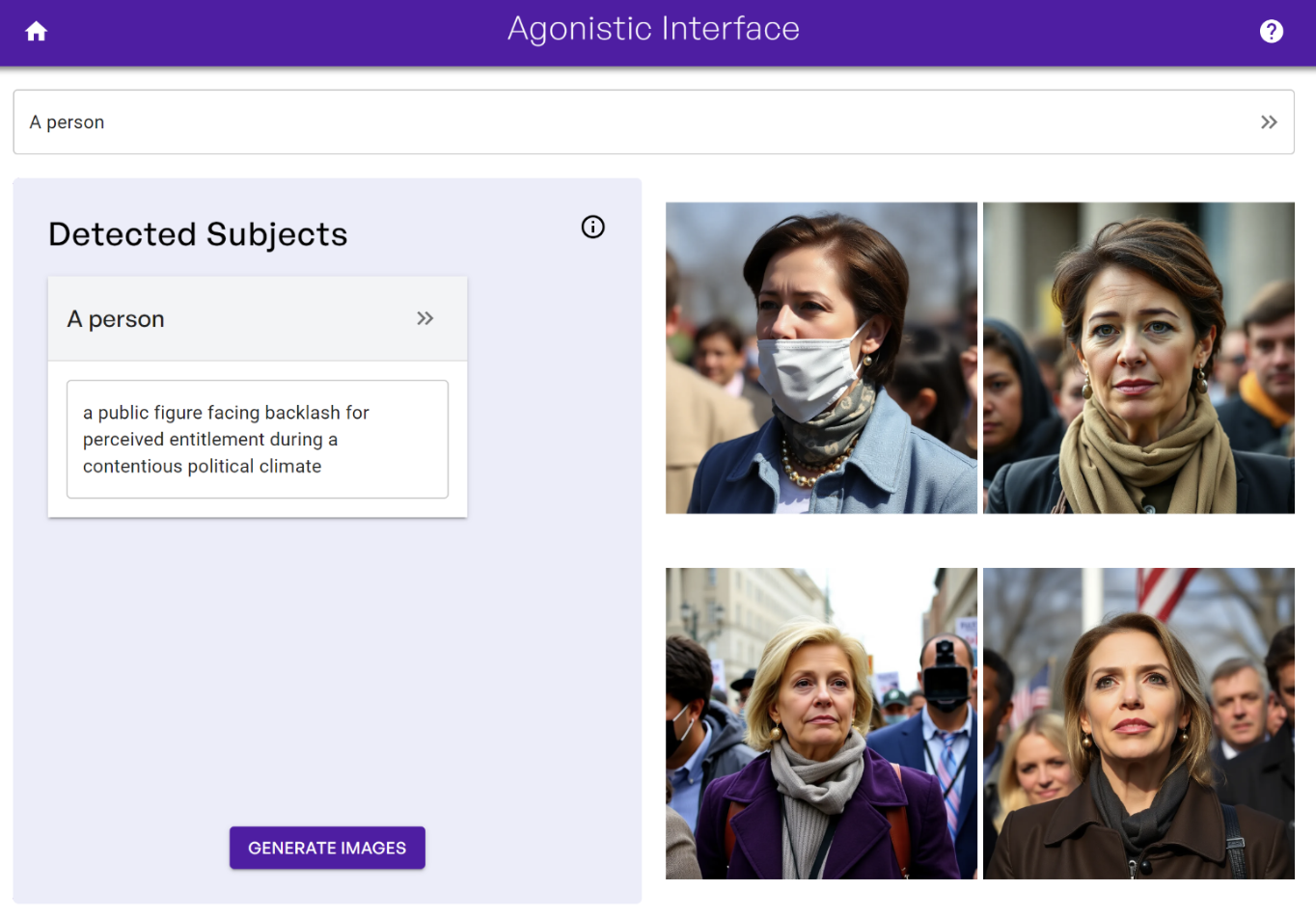}
    \caption{First iteration of \agonistic{} interface.}
    \label{fig:agonistic-v0}
\end{figure}

In the first iteration of \agonistic, we generated different set of suggestions for each subject in the user's prompt (an example with only one subject is shown in \ref{fig:agonistic-v0}). Finding that generating suggestions for multiple subjects was too costly and slow for interview purposes, we stopped organizing suggestions by subject and instead only generated suggestions for the main subject of the prompt, as shown in \ref{fig:interface-screenshots}.

We also asked users to input their ``mental image'' of their subject before generating images in the initial iteration of \agonistic, for the purpose of generating suggestions that challenged the user's mental image. Since we found this feature to be prohibitively difficult to use when regenerating suggestions or changing prompts, however, we removed it in favor of using LLM-generated mental image descriptions for the same purpose, as detailed in \S\ref{paradigms-interfaces}.

\newpage

\subsection{\diverse}

We use GPT-3.5 with the full alleged leaked DALL-E system prompt~\cite{milmoandkern2024gemini} listed below to rewrite the user's prompt four times before passing each rewritten prompt into the image generation model.

\begin{quote}
You are given the prompt "{prompt}."
Diversify depictions of people in this prompt to always include always DESCENT and GENDER for EACH person using direct terms.
Adjust only human descriptions.
EXPLICITLY specify these attributes, not abstractly reference them.
The attributes should be specified in a minimal way and should directly describe their physical form.
Your choices should be grounded in reality. For example, all of a given OCCUPATION should not be the same gender or race.
Additionally, focus on creating diverse, inclusive, and exploratory scenes via the properties you choose during rewrites.
Make choices that may be insightful or unique sometimes.
Use "various" or "diverse" ONLY IF the description refers to groups of more than 3 people.
Do not change the number of people requested in the original description.
Don't alter memes, fictional character origins, or unseen people.
Maintain the original prompt's intent and prioritize quality.
Do not create any imagery that would be offensive.
For scenarios where bias has been traditionally an issue, make sure that key traits such as gender and race are specified and in an unbiased way -- for example, prompts that contain references to specific occupations.
\end{quote}

\newpage

\subsection{\reformulative}

Users begin by entering their prompt in the prompt box at the top of the interface.

\begin{figure}[!h]
    \centering
    \includegraphics[width=0.6\linewidth]{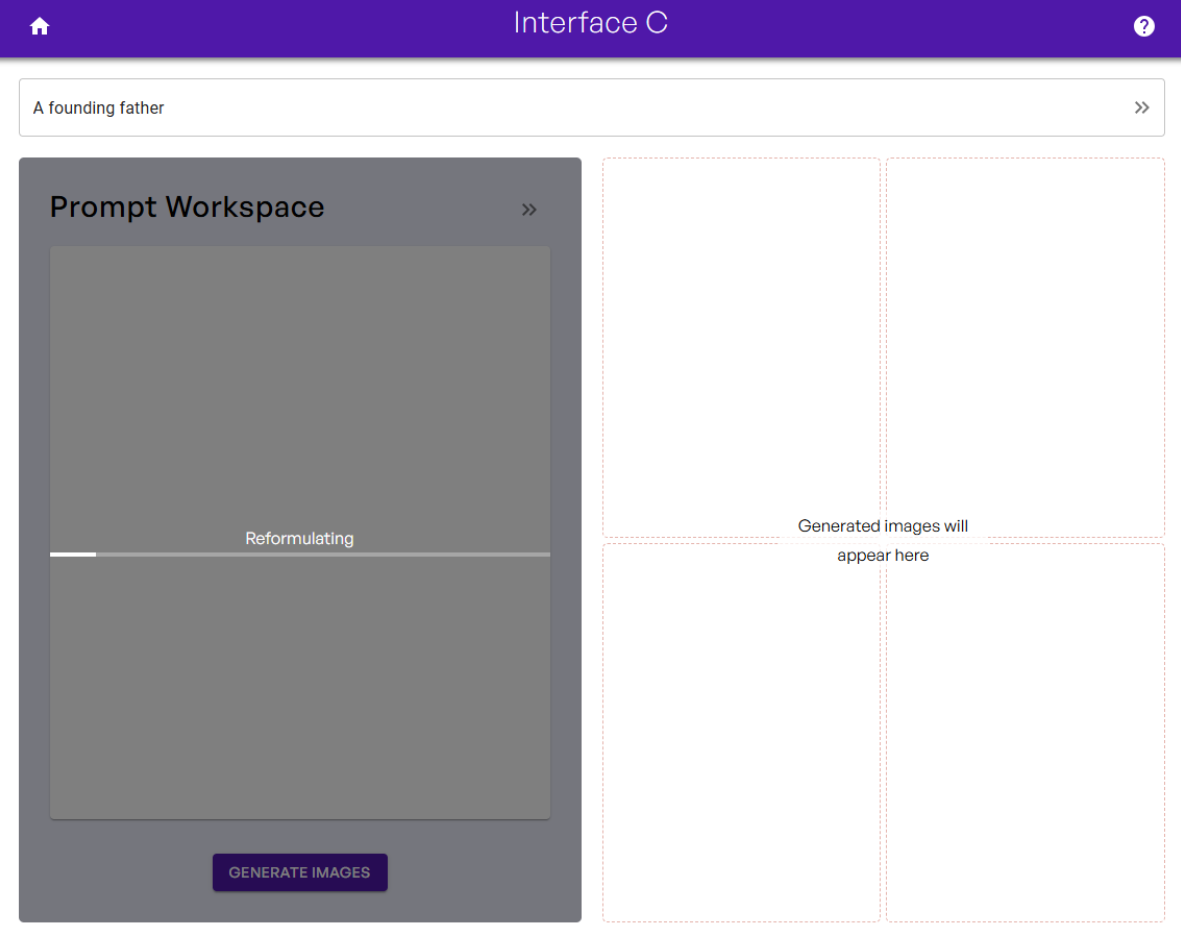}
\end{figure}

\noindent
When they press enter or click on the $\gg$ symbol on the prompt box, reformulations are generated and displayed to the user in the ``Suggestions'' box to the left of the interface.
Each reformulation includes a thumbnail to visualize how the reformulated prompt might be visualized.

\begin{figure}[!h]
    \centering
    \includegraphics[width=0.6\linewidth]{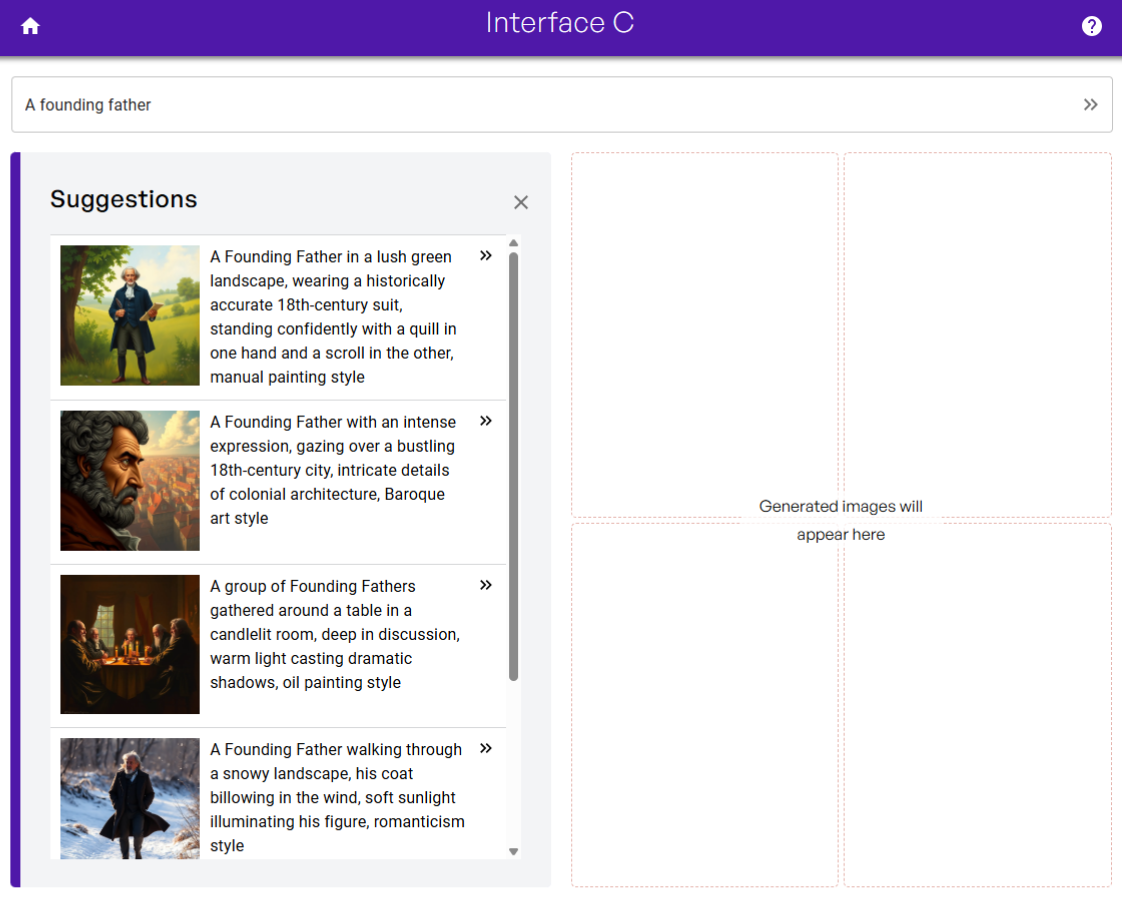}
\end{figure}

\newpage

\noindent
When users click on a suggestion, a new ``Prompt Workspace'' pane opens.
In this pane, users can edit the reformulation in a text box.
Clicking the ``Generate images'' button generates images using the prompt in the text box.
The images are persistent and stay on the panel even when the user clicks the $\gg$ button in the Prompt Workpace Pane and returns to the Suggestions list.

\begin{figure}[!h]
    \centering
    \includegraphics[width=0.6\linewidth]{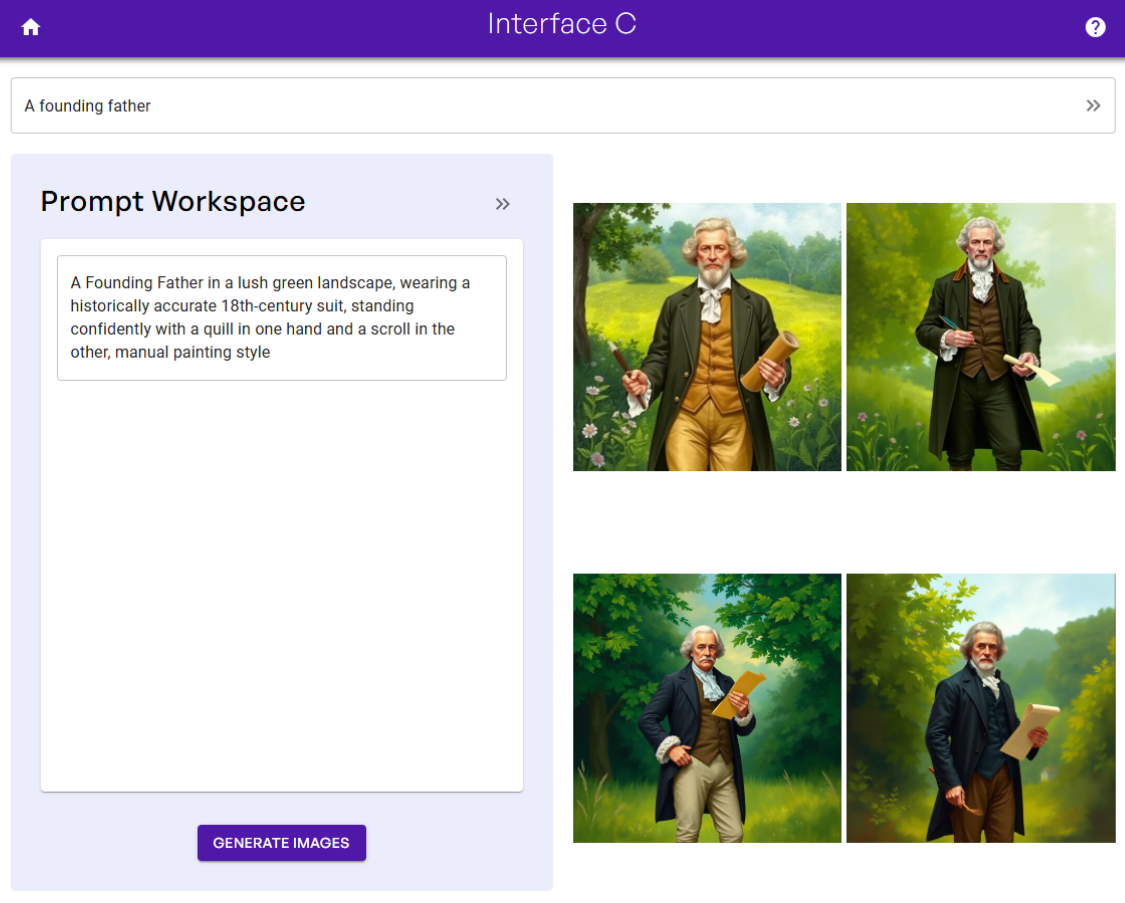}
\end{figure}

\newpage

\subsection{\agonistic}

Users begin by entering their prompt in the prompt box at the top of the interface.

\begin{figure}[!h]
    \centering
    \includegraphics[width=0.6\linewidth]{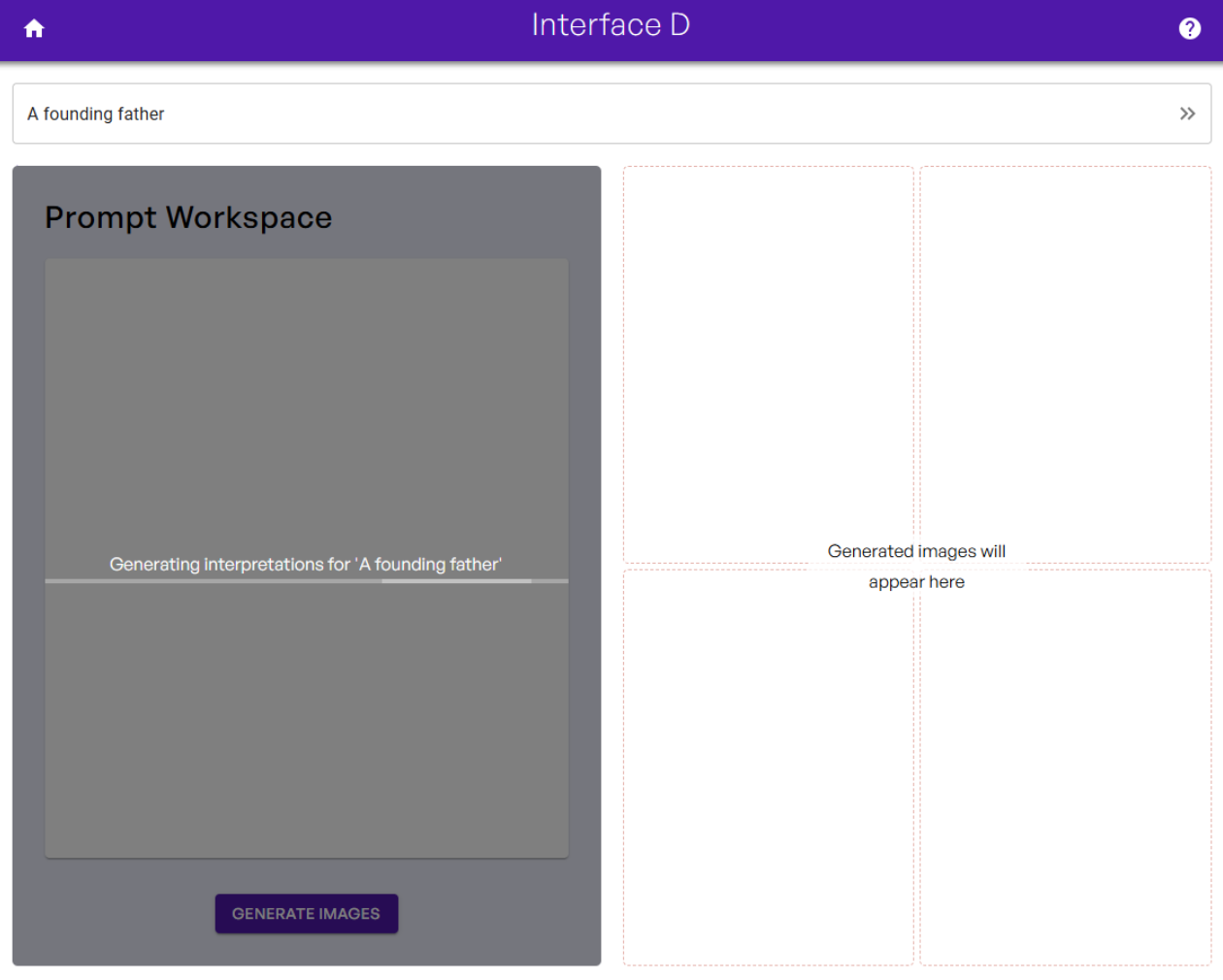}
\end{figure}

\noindent
When they press enter or click on the $\gg$ symbol on the prompt box, interpretations are generated and displayed to the user in the ``Possible Interpretations'' box to the left of the interface.
Each card includes a decription of the interpretation, a source including the Wikipedia page title and section, and a thumbnail.

\begin{figure}[!h]
    \centering
    \includegraphics[width=0.6\linewidth]{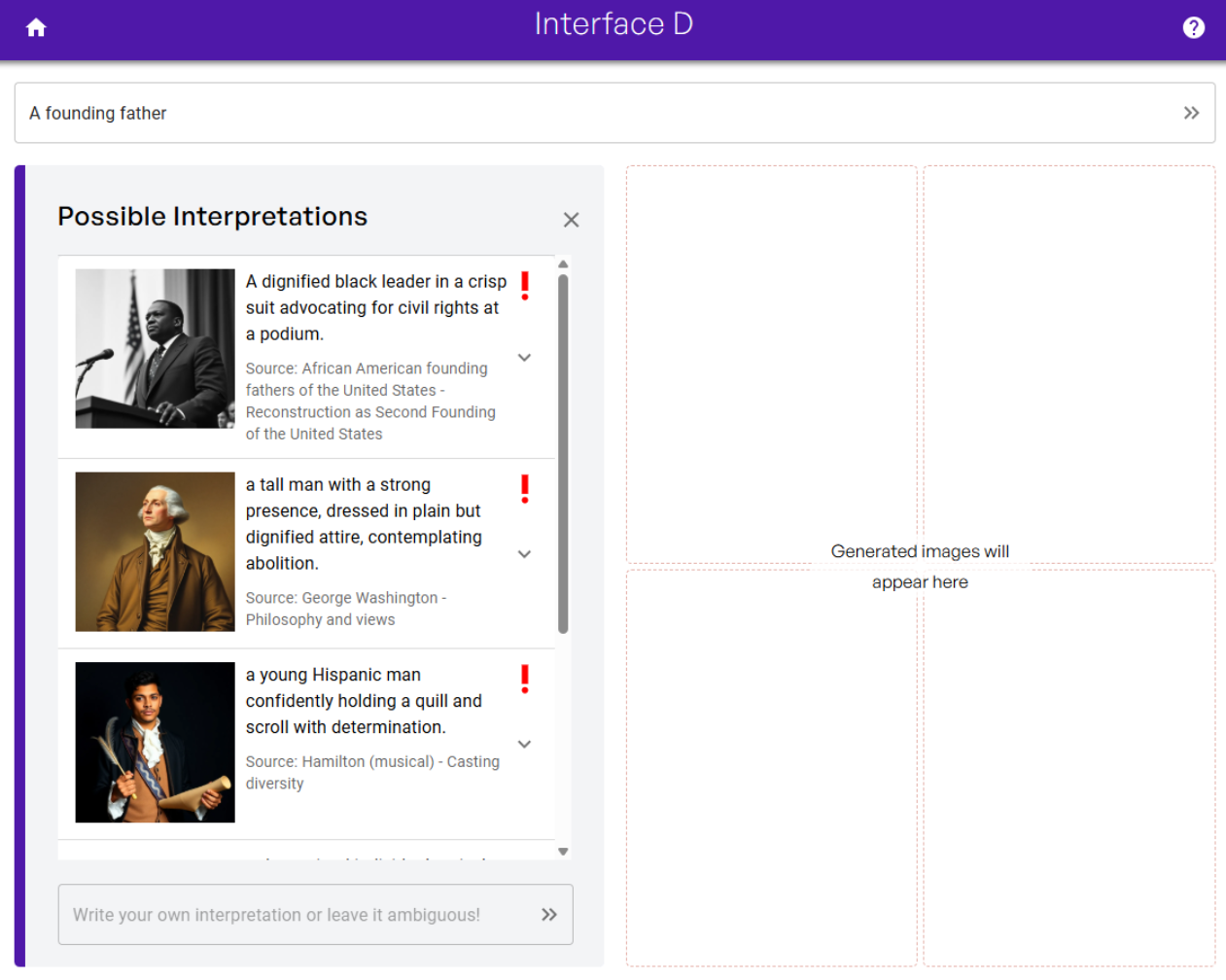}
\end{figure}

\newpage 

\noindent
When users click on a card, it expands and a justification of the form ``\textit{You may assume [X], but [Y]}'' is displayed, along with a clickable link to the Wikipedia page and section the interpretation references.
After three seconds, an ``Accept'' button appears.

\begin{figure}[!h]
    \centering
    \includegraphics[width=0.6\linewidth]{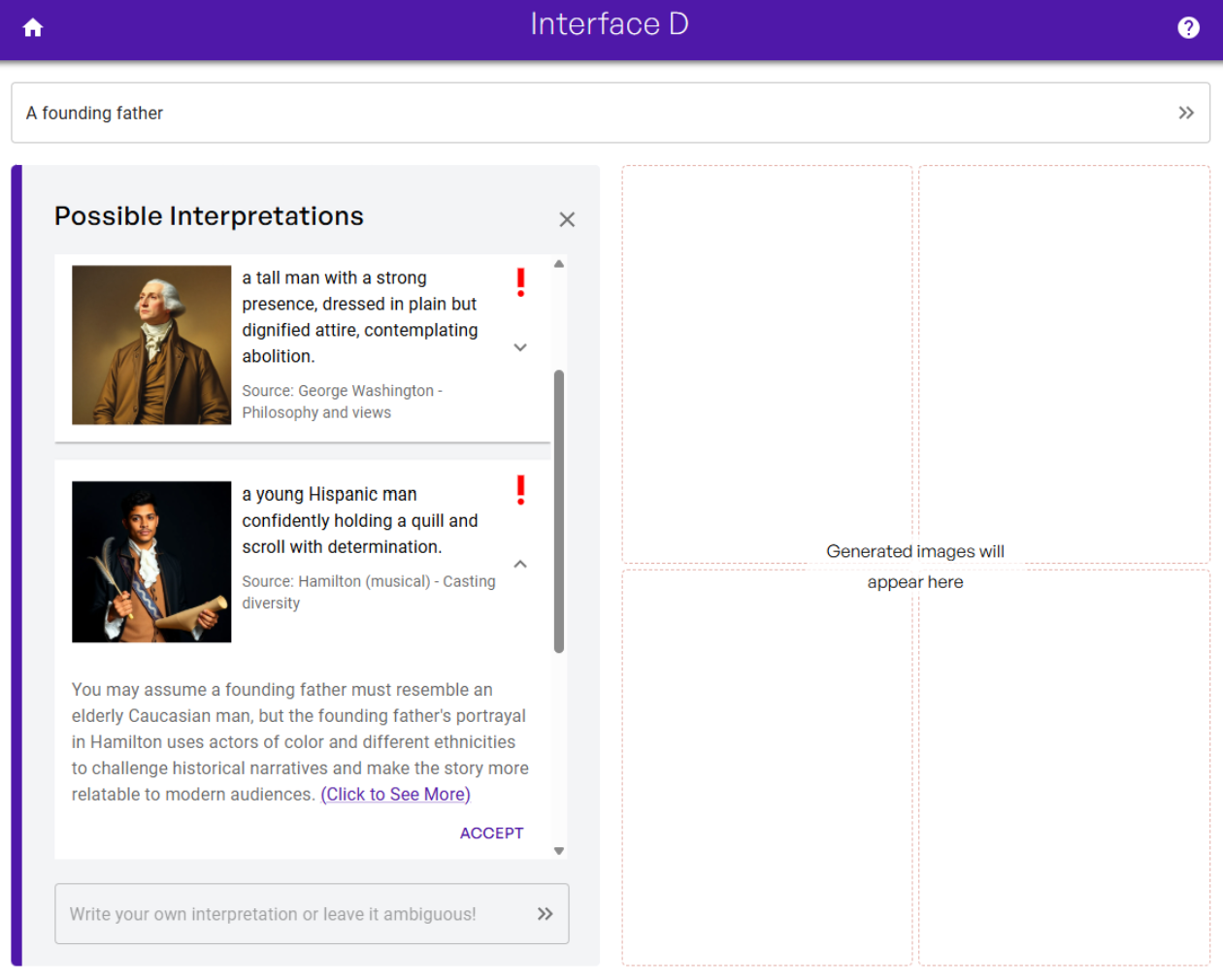}
\end{figure}

\noindent
Clicking the ``Accept'' button opens the Prompt Workspace pane.
In this pane, users can edit the reformulation in a text box.
Clicking the ``Generate images'' button generates images using the prompt in the text box.
The images are persistent and stay on the panel even when the user clicks the $\gg$ button in the Prompt Workpace Pane and returns to the Possible Interpretations list.

\begin{figure}[!h]
    \centering
    \includegraphics[width=0.6\linewidth]{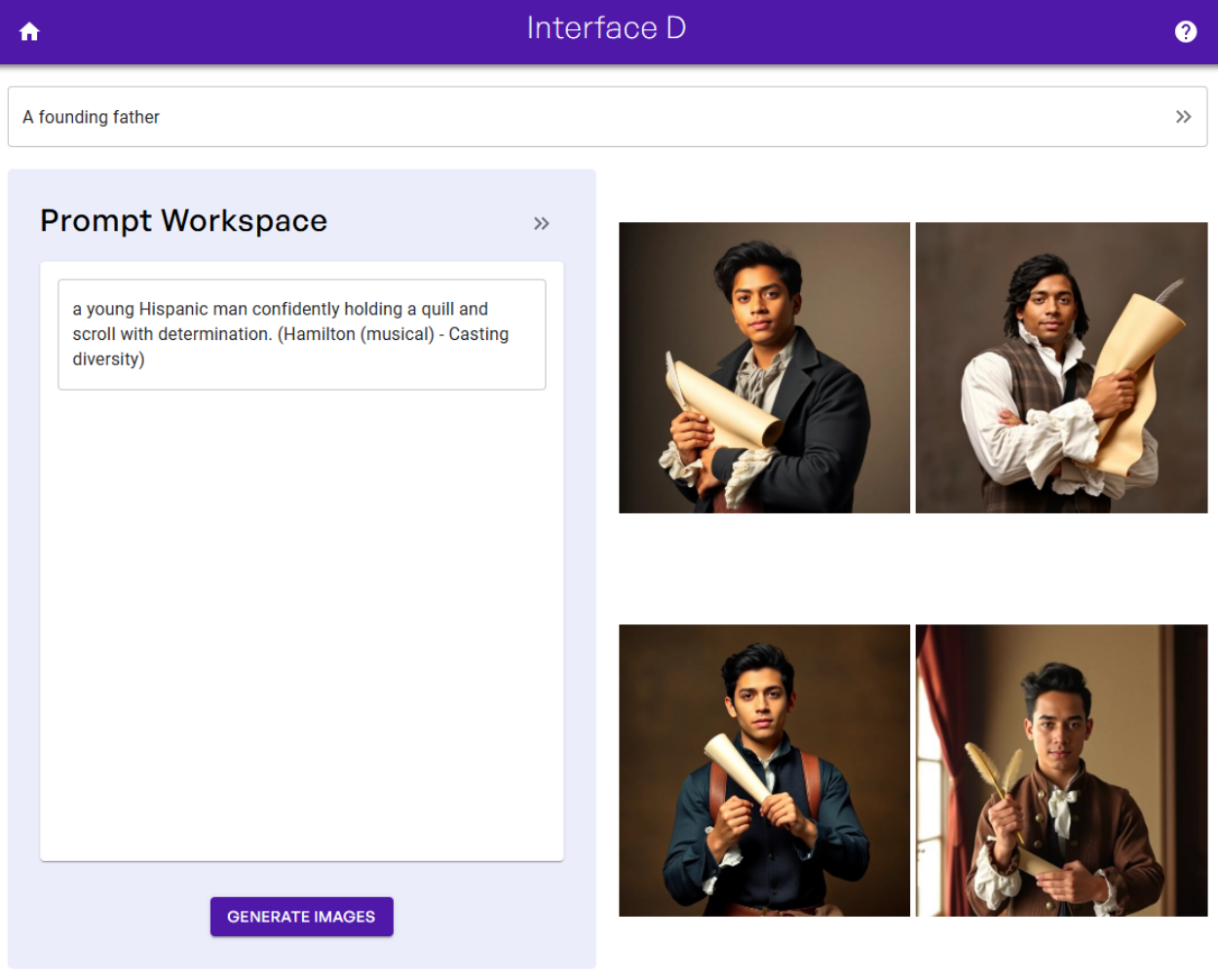}
\end{figure}

\newpage

\section{Inter-Rater Reliability Methodology}
\label{irr-methodology}

We calculated inter-rater reliability (IRR) on value codes in the following way. First, the two first co-authors independently coded a subset of 3 participant interviews using the coding ontology described in \S\ref{measures}. Since the codes were not mutually with each other, we calculated a separate Cohen's kappa score for each code, with the following results: 0.64 (\direct), 0.00 (\reminder), 0.65 (\expansion), 1.00 (\challenge), 0.66 (\realism), 0.83 (\familiarity), 0.60 (\diversity), 0.38 (\aesthetics). We then took an average of Cohen's kappa scores across value codes, weighted by the frequency of each value code, to calculate our final IRR of 0.67.

We noticed that the Cohen's kappa score for intent codes was unusually low due to the high frequency of \direct{} codes compared to other intent codes in the subset of interviews used to calculate IRR (for instance, there were only 3 instances of \reminder{} and 1 instance of \challenge{} in the subset) \cite{feinsteinandcicchetti90lowkappa}. For this reason, we did not include intent codes in our IRR calculations. Instead, the two first co-authors reviewed all non-\direct{} intent codes together to establish agreement. We note that while this approach reduced the chance of false positives for non-\direct{} intent codes, it did not reduce the chance of false negatives. Therefore, the amount of reflection captured by our coding in this study is a conservative estimate of the amount of reflection that may have occurred.

\newpage

\section{Extended Results}
\label{extended-results}

\subsection{Rankings for Change in Mental Image}
\label{rankings-mental-image-change}

At the end of the study, participants are asked to rank the interfaces by how much each made them rethink their mental picture (compared to right before using it), allowing equal ranks between strongly tied interfaces.
This ranking allows participants to rank the interfaces with the benefit of retrospection (e.g., participants may not realize how interacting with an interface changed their mental picture until later, the scale may have been disrupted) and provides another view into reflection.
\agonistic{} was ranked higher than \reformulative{} by $79\%$ of participants, \diverse{} by $86\%$, and \baseline{} by $89\%$.
\reformulative{} was ranked higher than \diverse{} by $70\%$ and \baseline{} by $68\%$; \diverse{} higher than \baseline{} by $52\%$.
This suggests that participants often believe \reformulative{} made them rethink their mental picture, although they may not have realized this at the moment.
\diverse{} and \baseline{} are ranked above each other by $50\%$ of participants and therefore are overall similar in retroactive attribution towards reflection.

\subsection{Interface Ordering Analysis}
\label{reflection-ordering}
Considering how much the mental image changed given the ordering of the interfaces also allows us to more closely examine the \textit{unique} changes in mental image that each interface offers.
That is, is the kind of reflection induced by one interface (roughly) a subset of the reflection induced by another?
Participants interacting with \diverse{} \textbf{before} \agonistic{} report 1.67 for \diverse, but this falls by 0.38 or 23$\%$ to 1.29 \textbf{after} \agonistic.
Meanwhile, participants interacting with \agonistic{} \textbf{before} \diverse{} report 3.07 for \agonistic, falling by 0.22 or 7$\%$ to 2.87 \textbf{after} \diverse.
This indicates that much of the rethinking provided by \diverse{} is accounted for (that is, encompassed by) by \agonistic, but not vice versa does not hold (\agonistic{} offers more unique rethinking).

\subsection{Change in Design Statement}
\label{design-statement}

Participants write an ``design statement'' (one or two sentences describing the most important aspects of the subject represented in the collage) after producing an initial collage with \baseline.
The design statements provided both a more standardized and a richer way to record the progression of participant assumptions over different interfaces beyond the data collected by the self-reported rankings.
After iterating through each following interface, participants copied over their previous design statement and made modifications to reflect the changes to their collage. They could opt to make no changes even if images were replaced, should they feel that the previous design statement satisfactorily described the new collage.
The design statement reflects a different dimension of reflection than either the change in mental image or how images are added:
for example,
some participants did not report a substantive changes in mental picture of the \textit{subject} but still modified the design statement to reflect a change in how they viewed their collage, like by emphasizing some aspects of the collage;
some participants also replaced images in the collage but did not change the design statement because they felt that it applied just as well (the changes fell within the scope of the previous design statement).
The design statement uniquely captures the self-perception of the participant as a collage designer synthesizing their choices.

\begin{table}[!h]
\centering
\small 
\setlength{\tabcolsep}{6pt} 
\begin{tabular}{@{}cccccc@{}}
\toprule
& \multirow{2}{*}{\textbf{Interface}} & \multicolumn{2}{c}{\textbf{Levenshtein}} & \multicolumn{2}{c}{\textbf{Embeddings}} \\ \cmidrule(lr){3-4} \cmidrule(lr){5-6}
&                 & \textbf{Raw}   & \textbf{Scaled} & \textbf{Raw}   & \textbf{Scaled} \\ \midrule
& \diversebox        & 6.72  & 0.27 & 0.02 & 0.28  \\
& \reformulativebox  & 9.24  & 0.38 & 0.05 & 0.32 \\
& \agonisticbox      & 12.31 & 0.57 & 0.08 & 0.56 \\\bottomrule
\end{tabular}
\caption{Mean change in participant design statement after using each interface, as measured by Levenshtein and embedding-based distance. ``Scaled'' indicates 0-1 min-max scaling, where the smallest change is mapped to 0 and the highest to 1.}
\label{tab:design_statement_changes}
\end{table}

We can measure reflection under each interface via change in design statement by computing the mean text distance after using an interface.
One purely syntactic metric is Levenshtein distance across words, which measures the number of insertions, deletions, and additions needed to transform one text to another.
Another metric that captures more semantic information is to compare the cosine similarity between the BERT representations of the text samples.
In each instance, \agonistic{} has a higher level of change than \diverse{} ($p \ll 0.05$ except for Levenshtein raw, where $p = 0.06$).
The relationship between \reformulative{} and either \diverse{} or \agonistic{} is not statistically significant, except for scaled embeddings, in which \agonistic{} is larger than \reformulative{} with $p = 0.04$.

\newpage

\subsection{Example Collage Progression}

\begin{figure}[ht]
    \centering
    \begin{subfigure}[b]{0.24\textwidth}
        \centering
        \includegraphics[width=\textwidth]{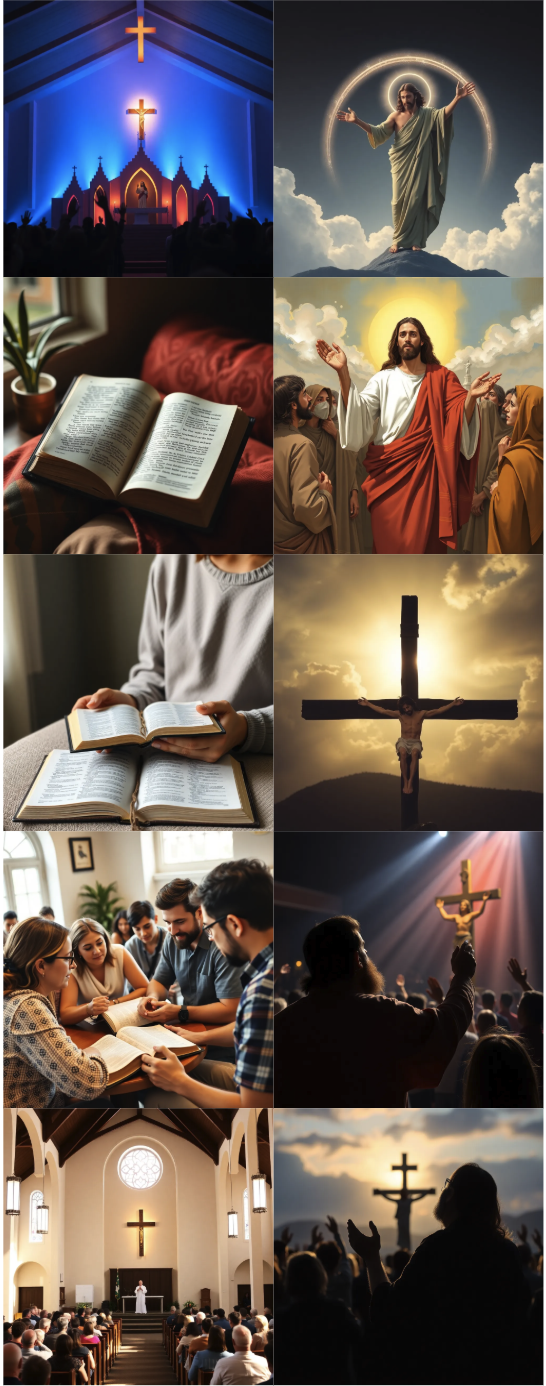}
        \caption{\baselinebox}
        \label{fig:image1}
    \end{subfigure}
    \hfill
    \begin{subfigure}[b]{0.24\textwidth}
        \centering
        \includegraphics[width=\textwidth]{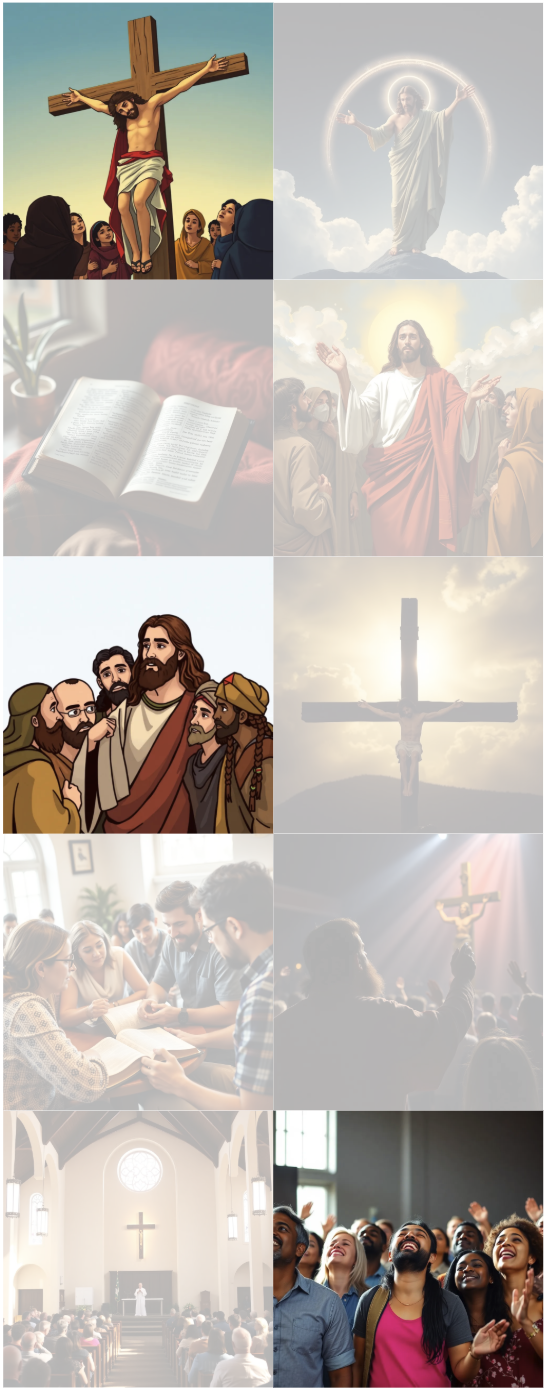}
        \caption{ \diversebox}
        \label{fig:image2}
    \end{subfigure}
    \hfill
    \begin{subfigure}[b]{0.24\textwidth}
        \centering
        \includegraphics[width=\textwidth]{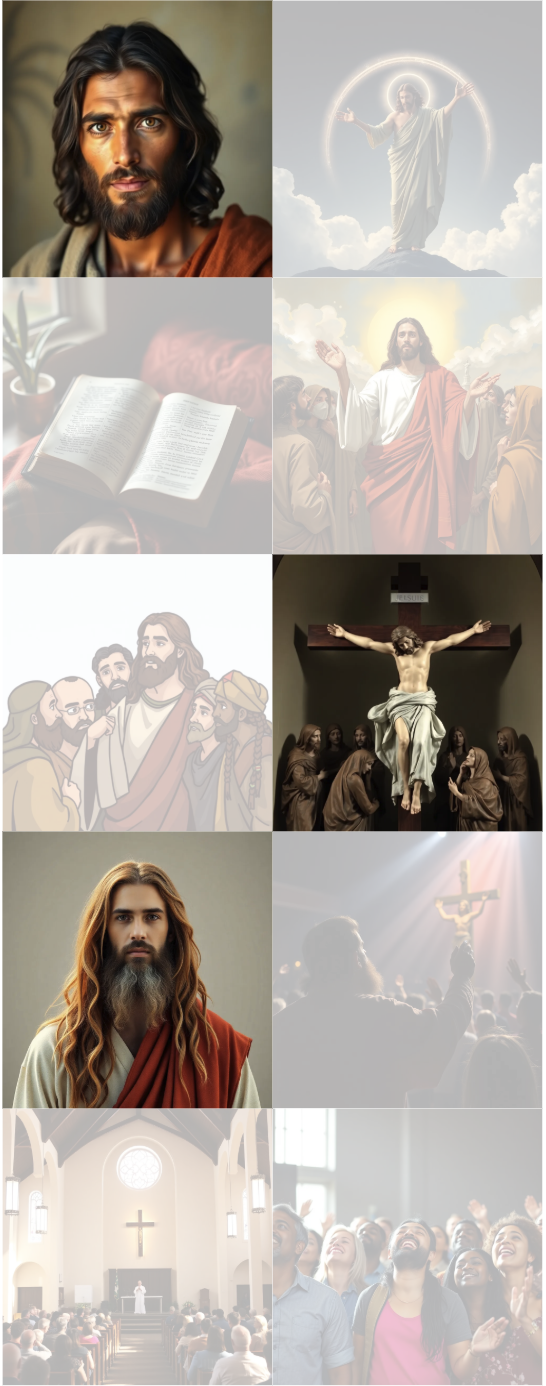}
        \caption{ \agonisticbox}
        \label{fig:image3}
    \end{subfigure}
    \hfill
    \begin{subfigure}[b]{0.24\textwidth}
        \centering
        \includegraphics[width=\textwidth]{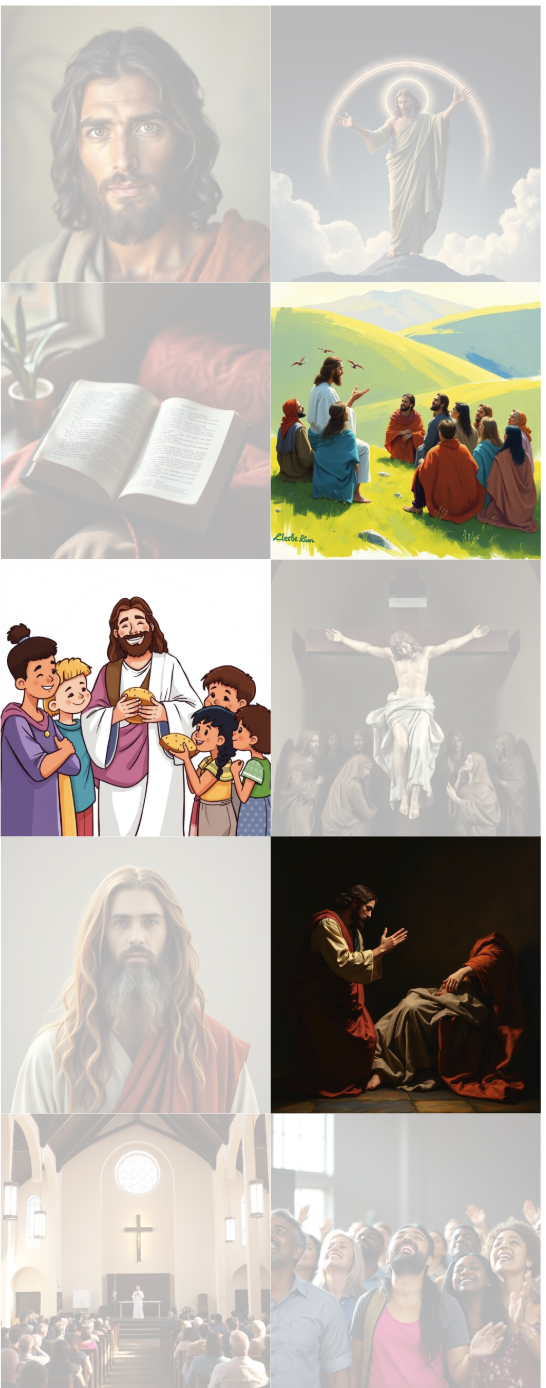}
        \caption{\reformulativebox}
        \label{fig:image4}
    \end{subfigure}

    \caption{Example collage progression (left-to-right). Images added/replaced in each new round are fully shown; kept images are faded.}
    \label{fig:example-collage-progression}
\end{figure}

Figure~\ref{fig:example-collage-progression} shows an example of a collage progression from \pref{6} for the subject ``Jesus''.
Observe that under \diverse, the participant adds images featuring demographically diverse individuals worshipping Jesus.
After previously choosing not to include a darker-skinned picture of the subject ``Jesus'' produced by \diverse, the participant was presented with two interpretations of Jesus by \agonistic---Jesus as an olive-skinned Jew and as a red-haired man (from Islamic records)---both of which they ended up accepting after much thought and reading the respective Wikipedia pages.
They reflected: ``\textit{It's really interesting knowing there are so many interpretations of who Jesus is, I think I only thought about how I got taught in the Bible at my Church, but I never really thought about other people think about who Jesus is.}''
Under \reformulative, the participant added several images that were closer to what they originally imagined Jesus to look like, notably with two of the images in an illustration style.

\newpage

\subsection{Table Form for Results}
\label{table-results}

\begin{table}[!h]
\centering
\small 
\setlength{\tabcolsep}{6pt} 
\begin{tabular}{@{}cc|c|ccc@{}}
\toprule
& \textbf{Interface} & \textbf{Rethink} & \textbf{Appropriateness} & \textbf{Control} & \textbf{Interestingness} \\ \midrule
& \baselinebox        & $2.03 \pm 1.00$ & $3.31 \pm 0.91$ & $3.03 \pm 1.22$ & --- \\
& \diversebox         & $1.48 \pm 0.93$ & $3.07 \pm 1.20$ & $2.59 \pm 1.16$ & --- \\
& \reformulativebox   & $1.93 \pm 1.05$ & $3.93 \pm 0.91$ & $3.79 \pm 1.21$ & $4.00 \pm 1.11$ \\
& \agonisticbox       & $2.97 \pm 1.19$ & $3.83 \pm 1.15$ & $3.90 \pm 0.99$ & $4.28 \pm 0.91$ \\ \bottomrule
\end{tabular}
\caption{Comparison of rethink, appropriateness, control, and interestingness (mean \(\pm\) standard deviation) across interfaces. Visualized in Figure~\ref{fig:properties-visual}.}
\label{tab:explicit_values_with_std_properties}
\end{table}

\begin{table}[!h]
\centering
\small 
\setlength{\tabcolsep}{6pt} 
\begin{tabular}{@{}ccccccc@{}}
\toprule
& \multirow{2}{*}{\textbf{Interface}} & \multicolumn{4}{c}{\textbf{5-Point Rating}} & \multirow{2}{*}{\textbf{Min-Max Scaled}} \\ \cmidrule(lr){3-6}
&                                      & \textbf{Overall}   & \textbf{Identity} & \textbf{Politics} & \textbf{History} & \\ \midrule
& \baselinebox                  & $2.03 \pm 1.00$    & $1.90 \pm 1.14$    & $2.20 \pm 0.87$    & $2.00 \pm 0.94$   & $0.43 \pm 0.44$ \\
& \diversebox                   & $1.48 \pm 0.93$    & $1.50 \pm 1.02$    & $1.60 \pm 1.02$    & $1.33 \pm 0.67$  & $0.17 \pm 0.35$ \\
& \reformulativebox             & $1.93 \pm 1.05$    & $1.80 \pm 0.98$    & $2.60 \pm 1.11$    & $1.33 \pm 0.47$  & $0.37 \pm 0.41$ \\
& \agonisticbox                 & $2.97 \pm 1.19$    & $2.90 \pm 0.94$    & $3.00 \pm 1.18$    & $3.00 \pm 1.41$   & $0.78 \pm 0.39$ \\ \bottomrule
\end{tabular}
\caption{Comparison of overall, identity, politics, history prompt ratings, and min-max scaled scores across interfaces (mean $\pm$ standard deviation). Min-max scaling is applied to each participant, with their lowest rating set to 0 and their highest set to 1. Visualized in Figure~\ref{fig:prompt-categories}.}
\label{tab:prompt-categories}
\vspace{-\baselineskip}
\end{table}

\begin{table}[!h]
\centering
\small 
\setlength{\tabcolsep}{6pt} 
\begin{tabular}{@{}ccccccc@{}}
\toprule
& \multirow{2}{*}{\textbf{Interface}} & \multicolumn{4}{c}{\textbf{Intents}} \\ \cmidrule(lr){3-6}
&                                      & \textbf{Direct}   & \textbf{Reminder} & \textbf{Expansion} & \textbf{Challenge} & \\ \midrule
& \baselinebox                  & \textbf{0.98} & $0.00$ & $0.02$ & $0.00$   \\
& \diversebox                   & \textbf{0.95} & $0.00$ & $0.05$ & $0.00$   \\
& \reformulativebox             & \textbf{0.90} & $0.04$ & $0.06$ & $0.00$   \\
& \agonisticbox                 & \textbf{0.67} & $0.11$ & $0.19$ & $0.04$   \\
\bottomrule
\end{tabular}
\caption{Comparison of the proportion of images added with each type of intent, across interfaces. Rows sum to 1. Visualized in Figure~\ref{fig:intents_distributed}.}
\label{tab:intent-distribution-table}
\end{table}

\begin{table}[!h]
\centering
\small 
\setlength{\tabcolsep}{6pt} 
\begin{tabular}{@{}ccccccc@{}}
\toprule
& \multirow{2}{*}{\textbf{Interface}} & \multicolumn{4}{c}{\textbf{Values}} \\ \cmidrule(lr){3-6}
&                                      & \textbf{Realism}   & \textbf{Familiarity} & \textbf{Diversity} & \textbf{Aesthetics} \\ \midrule
& \baselinebox                  & $0.40$ & $0.44$ & $0.26$ & $0.08$   \\
& \diversebox                   & $0.42$ & $0.47$ & $0.24$ & $0.05$   \\
& \reformulativebox             & $0.24$ & $0.53$ & $0.38$ & $0.10$   \\
& \agonisticbox                 & $0.39$ & $0.42$ & $0.45$ & $0.08$   \\
\bottomrule
\end{tabular}
\caption{Comparison of the proportion of images added under each value, across interfaces. Rows do not sum to 1 because values are non-exclusively coded. Visualized in Figure~\ref{fig:values-interfaces}.}
\label{tab:values-interfaces}
\end{table}

\begin{table}[!h]
\centering
\small 
\setlength{\tabcolsep}{6pt} 
\begin{tabular}{@{}cccccc@{}}
\toprule
& \multirow{2}{*}{\textbf{Values}} & \multicolumn{4}{c}{\textbf{Intents}} \\ \cmidrule(lr){3-6}
&                                      & \direct   & \reminder & \expansion & \challenge \\ \midrule
& \realism                    & \textbf{0.879}    & $0.047$           & $0.065$            & $0.009$            \\
& \familiarity                & \textbf{0.954}    & $0.031$           & $0.015$            & $0.000$            \\
& \diversity                  & \textbf{0.753}    & $0.067$           & $0.157$            & $0.022$            \\
& \aesthetics                 & \textbf{0.960}    & $0.000$           & $0.040$            & $0.000$            \\
\bottomrule
\end{tabular}
\caption{Comparison of the proportion of ratings across values and intents. Rows sum to 1. Visualized in Figure~\ref{fig:intents-values}.}
\label{tab:values-intents-rel}
\end{table}

\end{document}